\documentclass{article}
\usepackage{graphicx} % Required for inserting images
\usepackage{float}
\usepackage{natbib}
\usepackage{usebib}
\usepackage{comment}
\usepackage{authblk}
\usepackage[english]{babel}
\usepackage[title]{appendix}
\usepackage{tablefootnote}
\usepackage{geometry}
\usepackage{array}
\usepackage{adjustbox}

\usepackage{tikz}
\usepackage{xcolor}
\usepackage{framed}
\usetikzlibrary{positioning, arrows.meta, shapes.geometric}

\title{Breaking open the black box of the production function: an agent-based model accounting for time in production processes}

\author[1]{Jack Birner}
\author[2]{Marco Mazzoli}
\author[3]{Eleonora Priori}
\author[4]{Pietro Terna}
\affil[1]{University of Trento, Italia}
\affil[2]{University of Genova, Italia}
\affil[3]{School of Advanced Studies Sant'Anna, Italia}
\affil[4]{University of Torino, Italia}

\date{April 2024}

\begin{document}

\maketitle

\begin{abstract}
Traditional notions of production function do not consider the time dimension, appearing thus timeless and instantaneous. We propose an agent-based model accounting for the whole production side of the economy to unfold the production process from its very beginning, when firms receive production orders, to the delivery of the products to the market. In the model we analyze  with a high-degree of details how heterogeneous firms, having labor and capital as productive factors, behave along all the realization processes of their outputs. The main focus covers: i) the heterogeneous duration of firms’ production processes, ii) the adaptive strategies they implement to adjust their choices, and iii) the possible failures which may occur due to the duration of the production. Our agent-based model is a controlled experiment: we use a virtual \emph{central planner} mechanism, which acts as the demand side of the economy, to observe which firm individual behaviors and aggregate macroeconomic outcomes emerge as a reply to its different behaviors in a \emph{ceteris  paribus} environment.  Our applied goal, then, is to discuss the role of industrial policy by modeling production processes in detail.
\end{abstract}

\emph{Keywords}: agent-based models, production theory, time, national accounts, comparative analysis

\emph{JEL}: D21, D22, E01, E27, O21, P51

%%%%%%%%%%%%%%%%%%%%%%%%%%%%%%%%%%%%%%%%%%%%%%%%%%%%%%%%%%%%%%%%%%%%%%%%%%%%%%%%%%%%%%%
\section{Introduction}

In this work, we propose a journey to Hayekian foundation of complexity economics.
The first step will be modeling time in production processes through an agent-based simulation.

We build an agent-based model to consider the production side of the economy with heterogeneous firms, with labor and capital as productive factors. The aim of this model is to account with a high-degree of details how firms behave along all the realization processes of their outputs. The main focus covers: (i) the heterogeneous duration of firms’ production processes, (ii) the adaptive strategies they implement to adjust their choices, and (iii) the possible failures which may occur during the production. 

The model presents firms of different sizes with productive processes of different duration, lasting one or more time units (\emph{e.g.}, we consider months). In this way, production accounts for productive factors, following recipes about production techniques. This also allows the possibility of modeling failures due to the duration of the processes. Failures arise from the idea of variable time preferences, which may occur as the individuals' final choices can differ from the initial expectations \citep{Hayek.2007}. The production activities in each firm wait in productive queues and are then accomplished in parallel. Orders can have a duration of many time units, and for each order, failures can arise at any time. 
The adjustment of the labor and the capital quantities in each firm operates at given intervals.

\medskip

This model is related uniquely to the productive side of the economy and the supply formation: our applied goal is to discuss the role of innovation and industrial policy coevolution by modeling production processes in detail. To account for the demand side of the market, we introduce a central-planner-like mechanism, which performs a threefold task: (i) it generates the production orders assigning them to the firms, acting \emph{de facto} as a centralized demand mechanism, (ii) it retires goods from the firms’ inventories, and (iii) it distributes them investment goods according to different criteria that we want to compare. A first naive example is an uninformed central planner, wich acts retiring products from firms’ inventories and randomly assigning investments in a \emph{unwise} way---remembering \cite{10.2307/43828055}---, to move gradually to a \emph{wise} and informed planner assigning strategically investment goods to firms \citep{Mazzuccato2015}.

\medskip

Our agent-based model is a controlled experiment using a \emph{virtual} central planner to introduce a set of different situations related to the production process in a \emph{ceteris paribus} environment.

With this model, we start comparing our frame to that of the neoclassical literature. To do so we analyze in a formal way the behavior of our agents, and we look at our step by step construction as an improvement provided by ABMs built in a stock-flow consistent perspective \citep{JOES:JOES12221}. Following the critique to the standard assumptions of production theory of \cite{dosi2016production} and \cite{dosi2023curves}, we propose a model to overcome the traditional notion of \emph{production function}, better-tuning how production processes unfold over time and which events may interfere with the latter. In this spirit, we look at firms' production both as adapting to the market structure and as driving endogenously the business cycle \citep{AcemogluAzart2020, pangallo2020synchronization}.

%%%%%%%%%%%%%%%%%%%%%%%%%%%%%%%%%%%%%%%%%%%%%%%%%%%%%%%%%%%%%%%%%%%%%%%%%%%%%%%%%%%%%%%
\subsection{From ABM to ABBUMM: foundations for an Agent Based Bottom Up Model of the Macroeconomy}

Traditional top-down approaches present agents as stylized representations of theory in the model, producing inevitable relevant distances between theoretical propositions and the agents' content: the same distance that appears between theory and the real world. Instead, our construction starts with a set of realistically created agents, from whose action and interaction a bottom-up macroeconomic framework emerges. Agents can be tuned with elements derived from the heritage of macroeconomic theory thinking. That way, influence also passes through the agents' behavior to the bottom-up macroeconomic construction. For this reason, we introduce ABBUMMs (Agent Based Bottom Up Models for Macroeconomics) as a new class of models.

\begin{enumerate}
\item In a methodological-individualist perspective, causal mechanisms are ontologically located both at the level of the behavior of individuals and at that of the interactions between them. ABBUMMs can manipulate these elements by, for instance, running series of simulations in each of which one element is changed. The comparison of the resulting different scenarios is the heuristic tool that we use in our agent structure.

\item ABBUMMs can create ``macroeconomic'' data in the form of various scenarios to which we can apply our explanatory efforts. 

\item ABBUMMs make possible to single out and investigate some important unsolved explanatory problems in macroeconomics. Our chosen points of departure in Hayek's economics and methodology seem to be a promising path to follow. 

\end{enumerate}

Macroeconomics foundations represent a long search, considering too the bottom-up perspective \citep{gatti2011macroeconomics}. Our approach is quite innovative as we do not build the agent of the model as local equations derived from the macro relations, deriving calculations from that frame. Yet, we construct simple, realistic agents based on instances of classes or sets of types of agents. Then we observe the complex aggregate dynamics deriving from their actions, as we explain in Section \ref{model}.

%%%%%%%%%%%%%%%%%%%%%%%%%%%%%%%%%%%%%%%%%%%%%%%%%%%%%%%%%%%%%%%%%%%%%%%%%%%%%%%%%%%%%%%
\section{Introducing time in production: theoretical references}

In economic orthodox  approaches, the production function implements instantaneously the production process in time $t$. In this way the production function appears as timeless and instantaneous, and this assumption turns out to be very simplistic and unrealistic. An alternative theoretical framework has been provided by the work of of \cite{roegen1975model}.

Going beyond the unrealistic simplification of the timeless production process. This model accounts for the time sequence of the entrepreneurs planning and organizing all the activities ruling the production process from its very start when firms receive their production order to the end when the commodities are released on the market. Indeed, the model allows to use the framework not only for merely theoretical purposes but to exactly identify the origin of the events. Becker (1965) stresses the relevance of time in the production process although within the framework of the production function, explicitly formalizing it as an argument of the production process. However, it remained an under-explored topic in the economic theory despite its great potential.

Our approach, which explicitly accounts for the time elapsing within the production process, allows us to contemplate several events within the execution of the production. Thus, the model is thought to consider unplanned and unpredictable events that might alter the regular execution of the production process, \emph{e.g.}, sudden interruptions, unexpected internal conflicts or changes from the expectations due to varying time preferences. Specifically, the longer is the time elapsing of the production process the higher is the probability of such unplanned events.

%\medskip
%\textbf{\textit{Production processes are called \emph{processes} for a reason}}
%\medskip

%%%%%%%%%%%%%%%%%%%%%%%%%%%%%%%%%%%%%%%%%%%%%%%%%%%%%%%%%%%%%%%%%%%%%%%%%%%%%%%%%%%%%%%
\subsection{Production processes are called \emph{processes} for a reason: the production duration}

Macroeconomics uses highly aggregated models. Like all models, they make use of idealizations, \emph{i.e.}, they abstract from factors that are thought to be unnecessary for explaining the main causal mechanisms in an economy. In order for these models to be connected to empirical reality, these idealizations must be factualized (see \citealt{birner2001cambridge}).
However, if a model abstracts from---that is, excludes---factors that play a fundamental role in causal processes, this turns out to be an invitation for problems. Now, that is exactly what happened in the way in which production is modelled in macroeconomics, with production processes being described with the help of the production function, $ P=f(K,L)$. 
The implicit presupposition of the production function approach is that all variables have the same time index, so $ P_t=f(K_t,L_t)$. This function stands at the most for a correlation and not a process. Even in case the production function were $P_t=f(K_{t-1},L_t)$, for instance, it would stand for a correlation between output and capital in different periods, and at the most suggests that a process is involved.

The article ``Rehabilitation of Time Dimension of Investment in Macroeconomic Analysis'' by \cite{tsiang1949} is interesting for our topic for various reasons: it attributes the neglect of the period of production to three factors (pp. 204-5). First, the discovery that the period of production cannot be measured exactly. Second, the fact that the concept of the period of production was static. As a consequence, ``the baby [of the time dimension] is cast away with the bath water'' of the period of production (p.204). Thus, Keynes neglected time in his \emph{General Theory}. And third, the influence of Keynes drove other economists to neglect time, too. Hicks' attempt to draw attention to the importance of the role of time did not meet success. 

In the Hayek's view, if we want to express processes in a mathematical form, we have to use differences or differential equations. An alternative consists of diagrams in which time is one of the dimensions. Those are what Hayek used in \emph{The Pure Theory of Capital} \citeyearpar{hayek2007pure}. For us, the question is if the simulations with ABM depict or approximate sufficiently accurately these or similar diagrams. We think the answer is affirmative. Curiously enough, this is consistent with Hayek’s own view  \citep{hayek19812012}, in which he speaks of production processes in terms of rivers and their tributaries: to Don Lavoie (personal communication) he had expressed his interest in computer simulations of production processes.\label{Don}

The problem of a time-realistic production process has disappeared in recent years, with a hole in recent economic literature about the subject. Stop in a research path is not infrequent in science when the analysis path is more and more complex, with an insecure perspective of success. From the perspective of practical applications, the field is instead covered by business administration research using simulation and agent-based techniques.

Some examples are the supply chain management with the scheduling problem \citep{supplyChains2006}; the analysis of value flow in industrial production via simulation \citep{parv2019}; the analysis of the industrial production in an event-driven perspective 
\citep{app10124343}.

%%%%%%%%%%%%%%%%%%%%%%%%%%%%%%%%%%%%%%%%%%%%%%%%%%%%%%%%%%%%%%%%%%%%%%%%%%%%%%%%%%%%%%%
\subsection{Criteria for investment decisions: preferences varying over time generating failures}\label{varying}

Irving Fisher proposes as the fundamental criterion for investment decisions the rate of return over costs, which he defines as the `hypothetical rate of interest which if used in calculating the present worth of [\ldots] two options compared would equalize them or their differences (cost and return) [\ldots]'' \citep[155]{fisher1930}.
In the case of a multitude of investment options, entrepreneurs choose the production process(es), whose Net Rate of Return is $i$. But in real life, \emph{when} final products come to market is of fundamental importance for the health and survival of an enterprise. This is particularly so for innovative products. So, expected market conditions for final products are part of the decision process. 

Contrary to the aggregated and timeless approach of mainstream macroeconomics, our approach explicitly contemplates the time that elapses when the production process is started and, once it is started, during the production process. Furthermore, the model takes into consideration unplanned and unpredictable events that might alter the regular execution of the production process, e.g., sudden interruptions, unexpected internal conflicts or changes in expectations. In particular, the longer is the time that elapses between the beginning of the production process and the availability of the final product, the higher is the probability of such unplanned events. This is the negative counterpart of the higher productivity of longer production processes. 

Investment in production processes and the processes themselves take time, and during this time the conditions that made entrepreneurs think they would be profitable may change. For instance, if the demand for the final product turns out to be less than expected, entrepreneurs may discover this when it is too late (or vice versa, consumers may discover that there is not enough supply to satisfy their planned future demand for consumption goods at the price they expected). The expectations on which agents operating on the supply or on the demand side had based their plans are falsified. Therefore, they will adapt their plans. But in the meantime, resources have been diverted from one branch of industry, or from the production for one moment in the future, to another.

This lack of inter-temporal coordination may create excesses and wage rises or shortages and unemployment for the period it takes individuals to adapt their production and consumption plans. This is crucial in the real part of Hayek’s business cycle theory.

%%%%%%%%%%%%%%%%%%%%%%%%%%%%%%%%%%%%%%%%%%%%%%%%%%%%%%%%%%%%%%%%%%%%%%%%%%%%%%%%%%%%%%%
\subsection{Why agents?}

Quoting a key paper of \cite{axtell2000agents}:
\begin{quote}
    (\ldots) A second, more commonplace usage of computational agent models arises when mathematical models can be written down but not completely solved. In this case the agent-based model can shed significant light on the solution structure, illustrate dynamical properties of the model, serve to test the dependence of results on parameters and assumptions, and be a source of counter-examples. Finally, there are important classes of problems for which writing down equations is not a useful activity. In such circumstances, resort to agent-based computational models may be the only way available to explore such processes systematically, and constitute a third distinct usage of such models.
\end{quote}

We are mainly in the third case (``Finally \ldots'') because agents act in an independent time-related way and not only in sequential order, but also in a parallel way, as shown in Section \ref{model} and with two feedback loops, related to productive capacity adaptation and investment good availability.

\medskip

As far as it concerns the standard notion of production function, we highlight that if the production flow is constant, the time is implicit in the integration path in (\ref{prodfunc_integral}), with the function $f(L,K)$, defining the production density $P_t$:

\begin{equation} \label{prodfunc_integral}
    P_{(t_1,t_2)}=\int\limits_{t_1}^{t_2}f(L,K)dx
\end{equation}
 given the labor density $L_t$ and capital density $K_t$ at the instant $t$.
 
\medskip

\begin{figure}[h]
\begin{framed}
\centering
\includegraphics[width=0.9\textwidth]{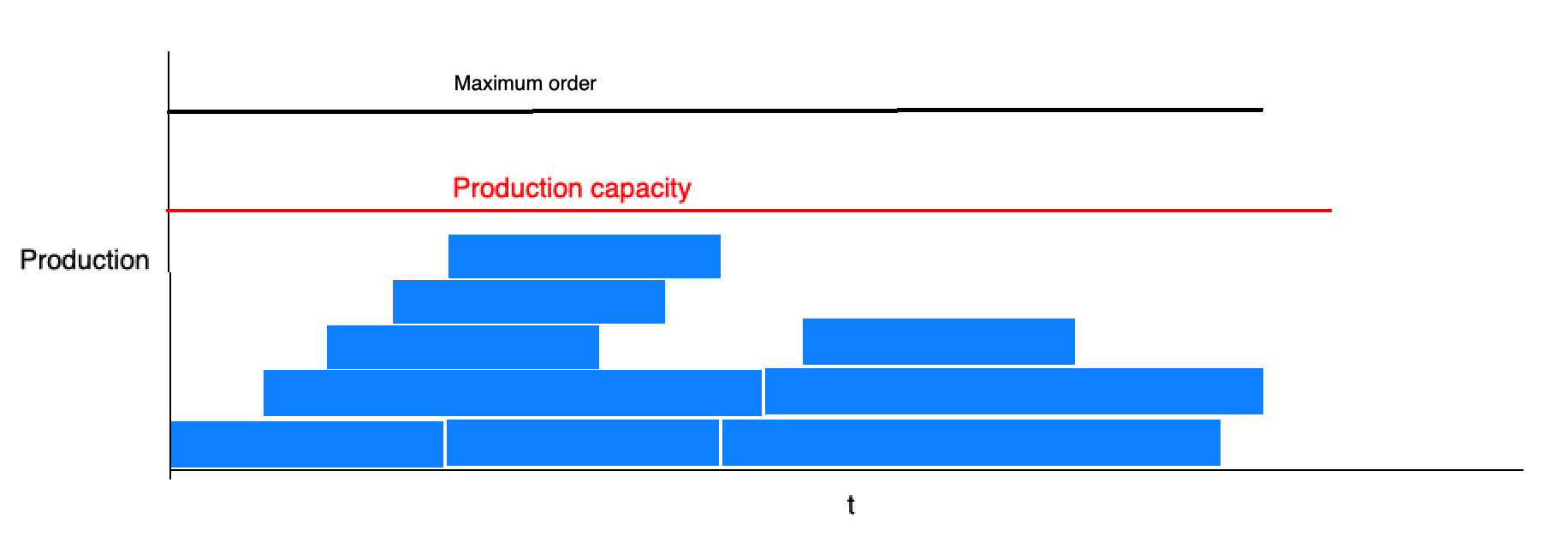}
\end{framed}
\caption{Random sequence of overlapping productions tasks with different durations}
\label{orders}
\end{figure}

Only ABMs allow us to manage a sequence of different duration orders as in Fig. \ref{orders}, where the production capacity (set in the model above the mean order and below the maximum one) limits the accepted orders and frequently produces unused $L$ and $K$.

Orders are generated in the interval $[s \cdot max, max)$ with $s>0$ as a share of the maximum value. This kind of analysis requires both the explicit duration of the production tasks and  the possibility of different durations. The different duration generates two effects. When it is higher: (i) it reduces unused production capability due to more frequent overlapping production tasks; (ii) it exposes to changes in preferences with wastes of production.

The emergent aggregate behaviors of Section \ref{results} come directly from this kind of construction.

%%%%%%%%%%%%%%%%%%%%%%%%%%%%%%%%%%%%%%%%%%%%%%%%%%%%%%%%%%%%%%%%%%%%%%%%%%%%%%%%%%%%%%%
\section{The model}\label{model}

We can imagine the world we build with this model as existing under a glass bell with random noise and a constant flow of orders but with a possible induced bias toward the production of consumption or investment goods. The latter are necessary to adapt and maintain the productive capacity, with strong feedback on the goods production.

The production mutually requires labor $L$ and capital $K$ (investments in durable, productive goods). $\frac{K}{L}$ fraction is the productive recipe specific for each firm. If global investments are limited, labor alone is useless. Firms are slightly undersized at the beginning of the simulation and find the right size in the start-up phase of events. Rationale: increasing capital is much easier than decreasing it. The labor and capital action and effects are reported in Appendix \ref{appLK}.

If a firm's total production capacity is insufficient, incoming orders are rejected; if only the temporarily unused capacity is, orders form a queue waiting for production. The long-term sequence of orders causes labor quantity adjustments, with hiring and firing actions, and capital quantity modification, with a complicated adaptation sequence, as shown in the Appendix \ref{appAdapt}.

%%%%%%%%%%%%%%%%%%%%%%%%%%%%%%%%%%%%%%%%%%%%%%%%%%%%%%%%%%%%%%%%%%%%%%%%%%%%%%%%%%%%%%%
\subsection{The event schedule in each time cycle}

The code of the simulation\footnote{\url{https://nbviewer.org/github/terna/ejmmp/blob/main/model1/model1.3.ipynb}} is built in Python using Repast4Py\footnote{\url{https://repast.github.io/repast4py.site/index.html}} as ABM framework for two reasons: its excellent scheduling tools and the capability of splitting the execution in several cores, maintaining whole interaction among the agents thanks the presence of \emph{ghosts} of the agents of other cores in each core.

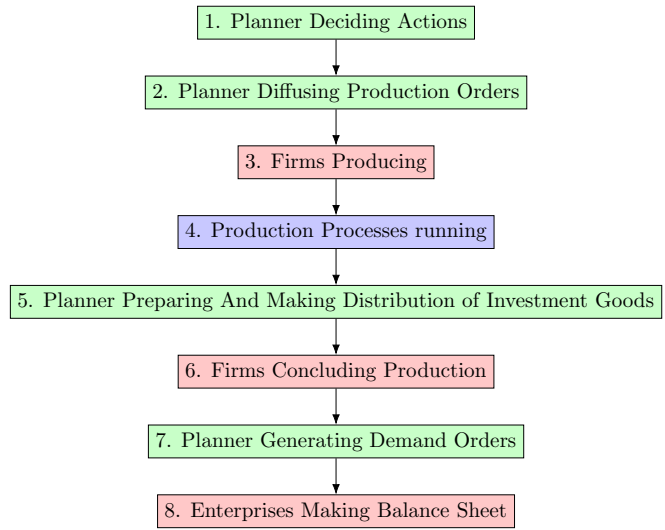
\begin{figure}[h]
\begin{center}
\scalebox{0.8}{

% Define colors
\definecolor{firmColor}{RGB}{255, 200, 200}
\definecolor{plannerColor}{RGB}{200, 255, 200}
\definecolor{processColor}{RGB}{200, 200, 255}

\begin{tikzpicture}[auto, node distance=0.6cm,>=Latex]
    % Nodes
    %\node[draw, fill=firmColor] (init) {Init Firms};
    \node[draw, fill=plannerColor] (plannerDeciding) {1. Planner Deciding Actions};
    \node[draw, fill=plannerColor, below=of plannerDeciding] (plannerDiffusing) {2. Planner Diffusing Production Orders};
    \node[draw, fill=firmColor, below=of plannerDiffusing] (firmsProducing) {3. Firms Producing};
    \node[draw, fill=processColor, below=of firmsProducing] (productionProcesses) {4. Production Processes running};
    \node[draw, fill=plannerColor, below=of productionProcesses] (plannerPreparing) {5. Planner Preparing And Making Distribution of Investment Goods};
    \node[draw, fill=firmColor, below=of plannerPreparing] (firmsConcluding) {6. Firms Concluding Production};
    \node[draw, fill=plannerColor, below=of firmsConcluding] (plannerGenerating) {7. Planner Generating Demand Orders};
    \node[draw, fill=firmColor, below=of plannerGenerating] (enterprisesBalancesheet) {8. Enterprises Making Balance Sheet};

    % Arrows
    %\draw[->] (init) -- (plannerDeciding);
    \draw[->] (plannerDeciding) -- (plannerDiffusing);
    \draw[->] (plannerDiffusing) -- (firmsProducing);
    \draw[->] (firmsProducing) -- (productionProcesses);
    \draw[->] (productionProcesses) -- (plannerPreparing);
    \draw[->] (plannerPreparing) -- (firmsConcluding);
    \draw[->] (firmsConcluding) -- (plannerGenerating);
    \draw[->] (plannerGenerating) -- (enterprisesBalancesheet);
\end{tikzpicture}

}
\caption{The sequence of the simulation steps}
\label{scheme}
\end{center}
\end{figure}

In Fig. \ref{scheme}, we introduce the simulation steps of each cycle over time. The simulation loop consists of scheduled repeating events that handle different aspects of the behavior of the agents:
    \begin{enumerate}
      \item The planner decides actions about production orders.
      \item The planner diffuses production orders.
      \item Firms accept orders if their productive capacity is sufficient: production involves checking resource availability, updating inventories, and managing productive processes.
      \item Productive processes run in parallel within each firm.
      \item The planner observes firms' demand for investment goods and decides distribution.
      \item Firms conclude production and make cost accounts.
      \item The planner generates demand orders for consumption and investment goods; the latter will be used in step 5 next time.
      \item Firms update their balance sheets with revenues, costs, and added value. Planner purchases are subject to random fluctuations, but as they relate to the value of each firm's finished goods inventory, they offset each other over time.
     \item National accounts and macroeconomic outcomes are observed by aggregating firm individual results.
    \end{enumerate}

The model has firms of different sizes with productive processes of
varying duration, lasting one or more time units (currently: months). 
The production activities are waiting in productive lines or queues and
then accomplished.
The model accounts for the details of the realization process, its duration, and the possible failures, analyzing the role of time in production processes.

%%%%%%%%%%%%%%%%%%%%%%%%%%%%%%%%%%%%%%%%%%%%%%%%%%%%%%%%%%%%%%%%%%%%%%%%%%%%%%%%%%%%%%%
\subsection{Production}

\begin{figure}[h]
\begin{center}
\scalebox{0.9}{

%\usepackage{tikz}
%\usetikzlibrary{shapes.geometric, arrows}

\tikzstyle{block} = [rectangle, draw, text width=1.5cm, text centered, rounded corners, fill=blue!20]
\tikzstyle{arrow} = [thick,->,>=stealth]

\begin{scriptsize}
\begin{tikzpicture}[node distance=2cm]

\node[block] (1) {productive\\process\\repository 1};
\node[block, right of=1] (2) {productive\\process\\repository 2};
\node[,right of=2](3){~};
\node[block, right of=3] (4) {productive\\process\\repository n};
\node[block, right of=4, yshift=1.5cm] (100) {central planner};

\node[block, below of=1, yshift=0cm,xshift=0.5cm] (5) {firm 1};
\node[block, below of=2, yshift=0cm,xshift=0.5cm] (6) {firm 2};
\node[,right of=6](7){\ldots};
\node[block, below of=4, yshift=0cm,xshift=0.5cm] (8) {firm n};
%\node[,right of=8](14){~};
\node[block, right of=8, yshift=0.6cm] (15) {each firm has its own balance-sheet};

%arrows describing links between firms and balancesheets (keep or drop?)
\draw [arrow] (5) |- (15);
\draw [arrow] (6) |- (15);
\draw [arrow] (8) |- (15);

\draw [arrow] ([xshift=-2cm]1) -> ([xshift=-2cm]5);
\draw [arrow] ([xshift=-1cm]1) -> ([xshift=-1cm]5);
\draw [arrow] (1) -> (5);
\draw [arrow] (2) -> (6);
\draw [arrow] ([xshift=-1cm]4) -> ([xshift=-1cm]8);
\draw [arrow] ([xshift=-2cm]4) -> ([xshift=-2cm]8);
\draw [arrow] ([xshift=-3cm]4) -> ([xshift=-3cm]8);
\draw [arrow] (4) -> (8);

\node[block, below of=5, yshift=0.5cm] (9) {L \& K\\adjustments\\in firm 1};
\node[block, below of=6, yshift=0.5cm] (10) {L \& K\\adjustments\\in firm 2};
\node[,right of=10, yshift=0.5cm](11){~};
\node[block, below of=8, yshift=0.5cm] (12) {L \& K\\adjustments\\in firm n};

\draw [arrow] (5) -> (9);
\draw [arrow] (6) -> (10);
\draw [arrow] (8) -> (12);

\draw [arrow] (9) |- + (-1,-1) |- (5); 
%the first -1 in (-1,-1) sets the horizontal size of the arrow
%shall we increase nodes distances to enlarge it a bit?
\draw [arrow] (10) |- + (-1,-1) |- (6);
\draw [arrow] (12) |- + (-1,-1) |- (8);

\draw [arrow] (100) -| (1);
\draw [arrow] (100) -| (2);
\draw [arrow] (100) -| (4);

\node[left of=1, yshift=1.5cm](44){t};
\node[below of=44,yshift=-4cm](45){t+1};
\draw [arrow] (44) -| + (-1,-1) |- (45);
\end{tikzpicture}
\end{scriptsize}

}

\caption{Firms with one or more production processes}
\label{firmsPPs}
\end{center}
\end{figure}
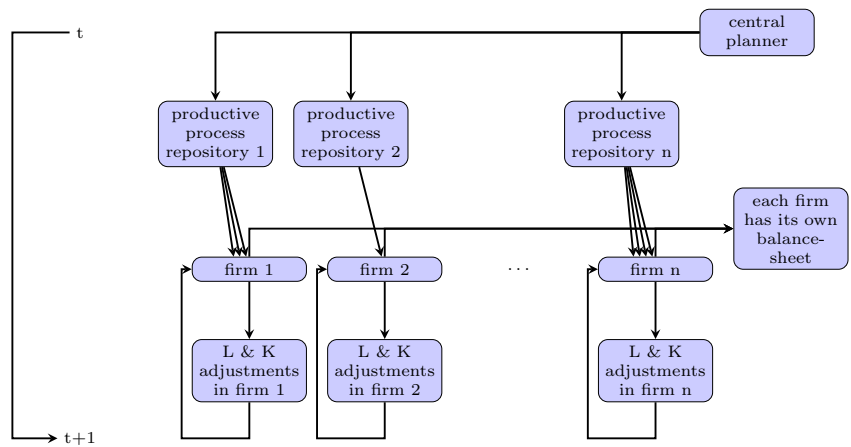

In Fig. \ref{firmsPPs} we explode the steps 3 and 4 of Fig. \ref{scheme}.

Firms receive orders as in step 3 above, accept them if they have sufficient total capacity and put them, aa future productive process, in the repositories represented in the upper part of Fig. \ref{firmsPPs}.

When production is possible, with free labor and capital, the productive processes run within each firm, possibly in a parallel way. The adjustment of the labor quantity L in each firm operates at given intervals, hiring and firing workers, following the previous sequence of orders, with their corresponding amounts. Capital adjustments follow, with some \emph{limitations}: we consider substitutions and increments of capital, corresponding to wished firms' investments, as described in Appendix \ref{appAdapt}.

Firms also account for failures in the production process, introduced in Section \ref{varying}. As consumption and productive goods mainly differ in the duration of the corresponding productive processes, the model implicitly allows the possibility of failures due to the
duration of the processes. The value of temporary inventories, related to orders under production for many units, will be corrected if a successive fail event arises. Failures arise from the Hayekian idea of variable time preferences,
which may occur as the agents’ final choices can differ from the initial expectations.

%%%%%%%%%%%%%%%%%%%%%%%%%%%%%%%%%%%%%%%%%%%%%%%%%%%%%%%%%%%%%%%%%%%%%%%%%%%%%%%%%%%%%%%
\section{Results}\label{results}

%%%%%%%%%%%%%%%%%%%%%%%%%%%%%%%%%%%%%%%%%%%%%%%%%%%%%%%%%%%%%%%%%%%%%%%%%%%%%%%%%%%%%%%
\subsection{Numerical assumptions}

The model uses version 1.3 of our code, with 10,000 firms and about 100,000 workers. Production orders are randomly generated and keep the same global amount in all the experiments.

To build up our firm population, we have selected a structure recalling Eurostat data,\footnote{\url{https://ec.europa.eu/eurostat/statistics-explained/index.php?title=Structural_business_statistics_overview#Size_class_analysis}} assuming that the maximum number of workers in a firm is set at 1000 and that  the firm dimension can be set according to its number of workers as follows:

1-9 workers $\to$ 93.7\%;
 
10-49 workers $\to$ 5.1\%;

50-249 workers $\to$  0.6\%;

$\ge$ 250  workers $\to$ 0.6\%.

Each firm produces consumption or investment goods according to the features assigned at its creation, regardless of size. The number of firms producing investment goods is significantly smaller than that of the firms producing consumption goods.

Table \ref{firm-feature} reports the characteristics of the firms of the simulation.

\begin{table}[ht]
\centering
\begin{adjustbox}{max width=0.8\textwidth}
\begin{tabular}{|>{\raggedright\arraybackslash}p{4cm}|*{8}{c|}}
\hline
\textbf{Descriptions} & \textbf{Class 1} & \textbf{Class 2} & \textbf{Class 3} & \textbf{Class 4} & \textbf{Class 5} & \textbf{Class 6} & \textbf{Class 7} & \textbf{Class 8} \\
\hline
Share of firms of each class & 0.843 & 0.094 & 0.034 & 0.017 & 0.003 & 0.003 & 0.003 & 0.003 \\
L min & 1 & 1 & 10 & 10 & 50 & 50 & 250 & 250 \\
L max & 9 & 9 & 49 & 49 & 249 & 249 & 1000 & 1000 \\
K min & 100 & 100 & 1200 & 1200 & 8000 & 8000 & 30000 & 30000 \\
K max & 450 & 450 & 2400 & 2400 & 16000 & 16000 & 70000 & 70000 \\
Order duration min & 1 & 2 & 1 & 2 & 2 & 4 & 6 & 12 \\
Order duration max & 1 & 4 & 1 & 4 & 4 & 8 & 12 & 24 \\
Recipe & 50 & 50 & 50 & 50 & 70 & 70 & 80 & 80 \\
L prod & 0.6 & 0.6 & 0.7 & 0.7 & 0.7 & 0.7 & 0.8 & 0.8 \\
Max order production & 6 & 6 & 50 & 50 & 250 & 250 & 500 & 500 \\
Assets' useful life & 12 & 12 & 12 & 12 & 12 & 12 & 12 & 12 \\
Planned markup & 0.10 & 0.10 & 0.30 & 0.30 & 0.20 & 0.20 & 0.30 & 0.30 \\
Order observation frequency min & 5 & 5 & 5 & 5 & 10 & 10 & 15 & 15 \\
Order observation frequency max & 10 & 10 & 10 & 10 & 15 & 15 & 20 & 20 \\
Production type & 0 & 1 & 0 & 1 & 0 & 1 & 0 & 1 \\
\hline
\end{tabular}
\end{adjustbox}
\caption{Parameters of firm classes}
\label{firm-feature}
\end{table}

Row names from 1 to 11 are self-explicating. In rows 12 and 13, we have the range of the frequency of past order observation to modify $L$ and $K$. The \emph{Production type} has $0$ value for firms producing consumption goods and $1$ for those making investment goods.

%%%%%%%%%%%%%%%%%%%%%%%%%%%%%%%%%%%%%%%%%%%%%%%%%%%%%%%%%%%%%%%%%%%%%%%%%%%%%%%%%%%%%%%
\subsection{The plan of the experiments}\label{expPlan}

First, we introduce a set of experiments useful to test and describe the model's behavior.
\begin{itemize}

\item The first three cases are just control tests to check the functioning of the model: cases \ref{Planner zero}, \ref{Planner total} and \ref{Planner random}, whether the planner assigns zero, the total quantity or a random fraction of the investment goods required by the firms (step 5 in Fig. \ref{scheme}). In the second and third cases, the planner operates regardless of the quantity of investment goods collected in step 7 of the same figure in the previous cycle.
\end{itemize}

The second block of three experiments shows the main results of our model. There, we observe the effects of implementing policies fostering the production orders for consumption or durable productive goods. 
Each of these three is run twice to observe how they change with different production durations: 
\begin{itemize}

\item a \emph{regular} production order policy, which does not introduce distorting mechanisms in the economy, with cases \ref{plannerPropReg1} and \ref{plannerPropReg2};

\item a \emph{pro-consumption} policy, which supports the sectors producing consumption goods, with cases \ref{PlannerPropMin1} and \ref{PlannerPropMin2};

\item  a \emph{pro-industry} policy, which supports the sectors producing durable productive goods, with cases \ref{PlannerPropMax1} and \ref{PlannerPropMax2}.
\end{itemize}

In each of these cases, the central planner assigns a quantity of investment goods that is a) proportional to that required by the firms and b) constrained by the quantity of investment goods bought by the planner itself in the previous cycle. When the central planner applies distorting mechanisms, the adjustments are fixed proportionally to the initial value of the consumption and investment goods production.

\begin{itemize}
\item 

The last of these scenarios is performed changing also another assumption: namely, increasing the parameter that regulates the probability of failures to observe changes due to the varying preferences (see Section \ref{varying}), in cases \ref{PlannerPropMax1Fail0.1} and \ref{PlannerPropMax2Fail0.1}.

\end{itemize}

%%%%%%%%%%%%%%%%%%%%%%%%%%%%%%%%%%%%%%%%%%%%%%%%%%%%%%%%%%%%%%%%%%%%%%%%%%%%%%%%%%%%%%%
\subsection{Order generation}\label{ordGen}

In the model and, most of all, in the real world, firms need to size their production capacity to the level of higher-value orders because, too frequently, production capacity would remain unused.
They do not even size it on the average order value; they would lose too many orders upward. Standard sizing is between the average and maximum levels, generating a quite high unused productive capacity. Of course, too frequent layoffs and hiring decisions are not realistic.

To avoid having a too wide distribution of orders, with the presence of minimal values, we introduce a positive minimum proportional to the maximum possible value for which orders lie in the range $[s \cdot max, max)$ with $s>0$ as a share of the maximum value.  

The value of orders can be increased in favor of the production of consumption goods or capital goods, in either case with a proportional reduction in the other component so that the total flow of orders remains constant.

%%%%%%%%%%%%%%%%%%%%%%%%%%%%%%%%%%%%%%%%%%%%%%%%%%%%%%%%%%%%%%%%%%%%%%%%%%%%%%%%%%%%%%%
\subsection{Building bridges between theory and model results}

A synthetic overview of the experiments:
\begin{itemize}

\item 
Comparing the results in \ref{plannerPropReg1} and in \ref{plannerPropReg2}, we observe the effect of doubling the duration of production processes under normal order conditions without distortion in favor of consumption or investment goods.

The GDP level at the end of the period is the same, but it is reached immediately in the second case, the one with a doubling duration; another difference is the level of consumption goods, which is much higher than investment goods if the production has a longer duration.

We observe the effect of an \emph{advantage in production capacity utilization}, which mainly affects the small firms, most of all devoted to consumption goods. As indicated at the beginning of Section \ref{ordGen}, firms size their production capacity above the average for individual orders. This situation applies to the model and reality. Frms also have to accept many orders that, in successions of time, are well below production capacity; if the duration increases, since they can process orders in parallel, production capacity utilization increases when small orders, which last as an example, two periods instead of one, are placed side by side overlapping.

The framework described determines the success of small firms, as many as in reality and dedicated to consumption goods; these firms tend to take investment away from those producing the capital goods, so they produce less than in the case of short production duration.

\item 

With the results in \ref{PlannerPropMin1} and in \ref{PlannerPropMin2}, we observe the effect of doubling the production time when we reduce the volume of orders directed to the production of capital goods while increasing that of consumption goods.

Firms producing consumer goods start at a massive advantage over those producing capital goods, an advantage much reinforced if the processes are twice as long, with a higher level of GDP in that case. In this case, however, the level of GDP is lower than in all other situations.

Those firms need a lot of productive capacity, which they do not find because few capital goods are produced; lacking substitution, $K$ falls, and all economic indicators fall.

Small firms, if the duration is short, when faced with larger orders, lose many orders, not having sufficient capacity; if the duration is double, the effect mentioned above is acting, at least helping in recovering the lower value orders in parallel, working more than one in parallel, with the \emph{advantage in production capacity utilization}.

\item 
With the results in \ref{PlannerPropMax1} and in \ref{PlannerPropMax2}, we observe the effect of doubling the production time when we increase the volume of orders directed to the production of capital goods while decreasing that of consumption goods.

In this case, however, the GDP level is lower than in all other situations. The overall output level is now much higher (investment effect), but growth is difficult because there is a need to greatly increase production capacity, with capital goods available but only gradually over time.

By doubling durations, consumer goods weigh more for the reason mentioned above (the many small firms that greatly increase their output in parallel). Still, the strong growth allowed by the duration effect in small firms and the super demand for investments restrain the growth of productive capacity in all firms.

\end{itemize}

%%%%%%%%%%%%%%%%%%%%%%%%%%%%%%%%%%%%%%%%%%%%%%%%%%%%%%%%%%%%%%%%%%%%%%%%%%%%%%%%%%%%%%%
\subsection{Simulation experiments}

In the following figures, we observe what happens to the planner's behavior and the yearly national accounts under these scenarios. For each of the following pictures, the first graph is expressed in terms of months as time units (also including two years of the model's technical \emph{warming-up}). In contrast, in the second graph, the time unit is converted in \emph{years}, and the \emph{warming-up} phase is not recorded. Remark: reality proceeds ``from always,'' our world begins when we start the simulation.

Please notice that when we refer to investments, we consider them a gross measure as they include both substitutions and increments of productive durable capital goods.

%%%%%%%%%%%%%%%%%%%%%%%%%%%%%%%%%%%%%%%%%%%%%%%%%%%%%%%%%%%%%%%%%%%%%%%%%%%%%%%%%%%%%%%
\subsubsection{Planner zero}\label{Planner zero}

\begin{figure}[H]
\begin{center}
    \includegraphics[width=0.7\textwidth]{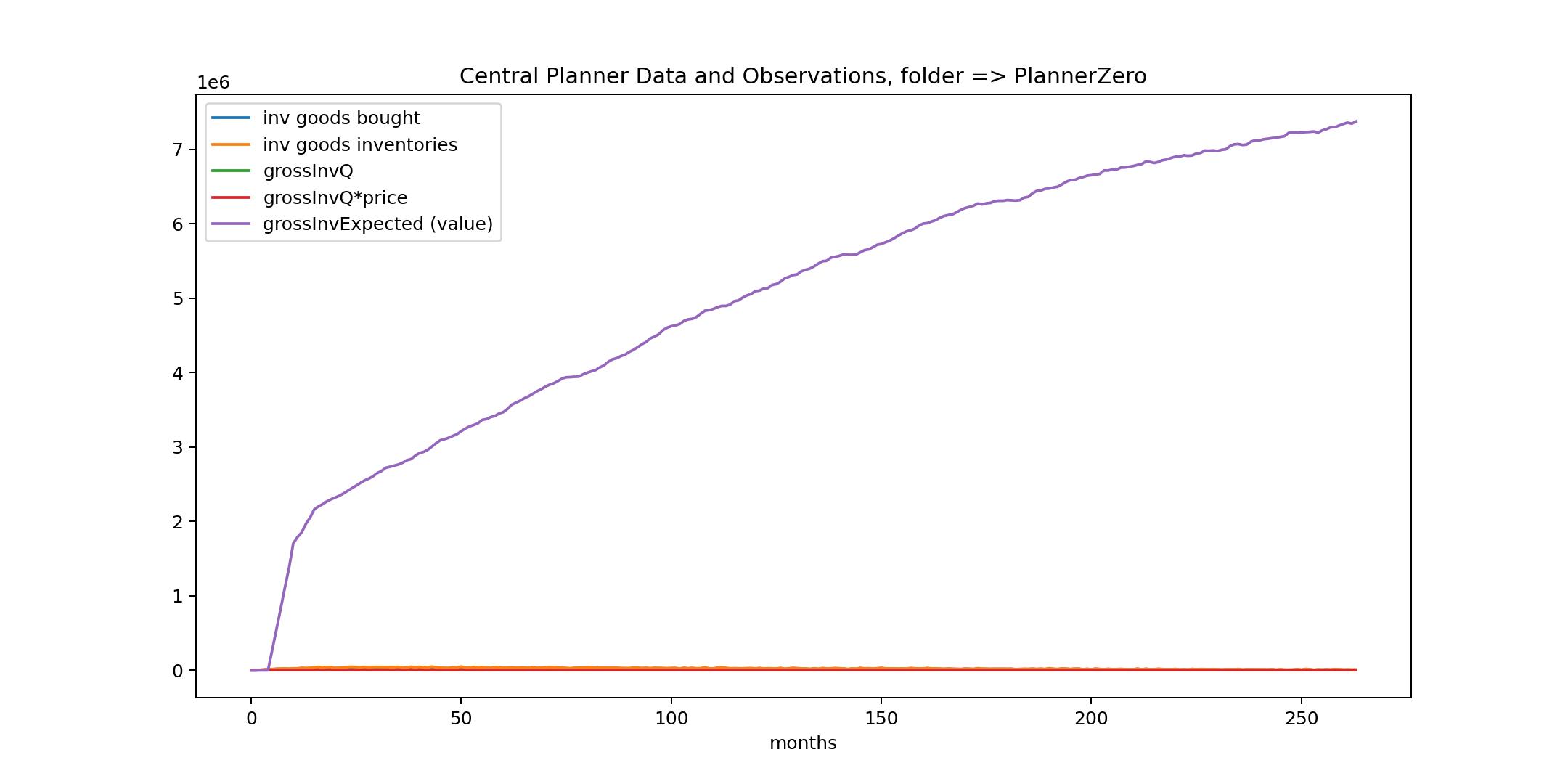}
\end{center}
\caption{The planner assigns zero of the investment goods required by the firms}
\label{PlannerZeroFig}
\end{figure}

We start with a control run to verify the model's behavior under stress.

In Fig.~\ref{PlannerZeroFig}, the planner behavior plot shows what happens when the latter assigns zero to the investment goods that the firms require: as they do not receive any of their requests, the amount of what they desire \big(\emph{grossInvExpected (value)}\big) experiences, after an initial boost, a constant growth at a slower but still sustained pace.

The other series that we introduce here also for the successive cases are: 
\begin{itemize}\label{seriesExplanation}

\item \emph{inv goods bought}, as investment goods bought by the planner (step 5 of Fig. \ref{scheme}); in the case of the \emph{planner zero} they go rapidly to zero as, missing the substitutions due to planner's action, the production of investment goods declines;

\item \emph{inv goods inventories} are the inventories of investment goods left after the central planner's previous step; they can be bought in successive time units;

\item \emph{grossInvQ} is the gross investment of the firms, deriving from the action of the planner always in step 5, in quantity;

\item \emph{grossinvQ*price}, the same in value.

\end{itemize}

\begin{figure}[H]
\begin{center}
\includegraphics[width=0.9\textwidth]{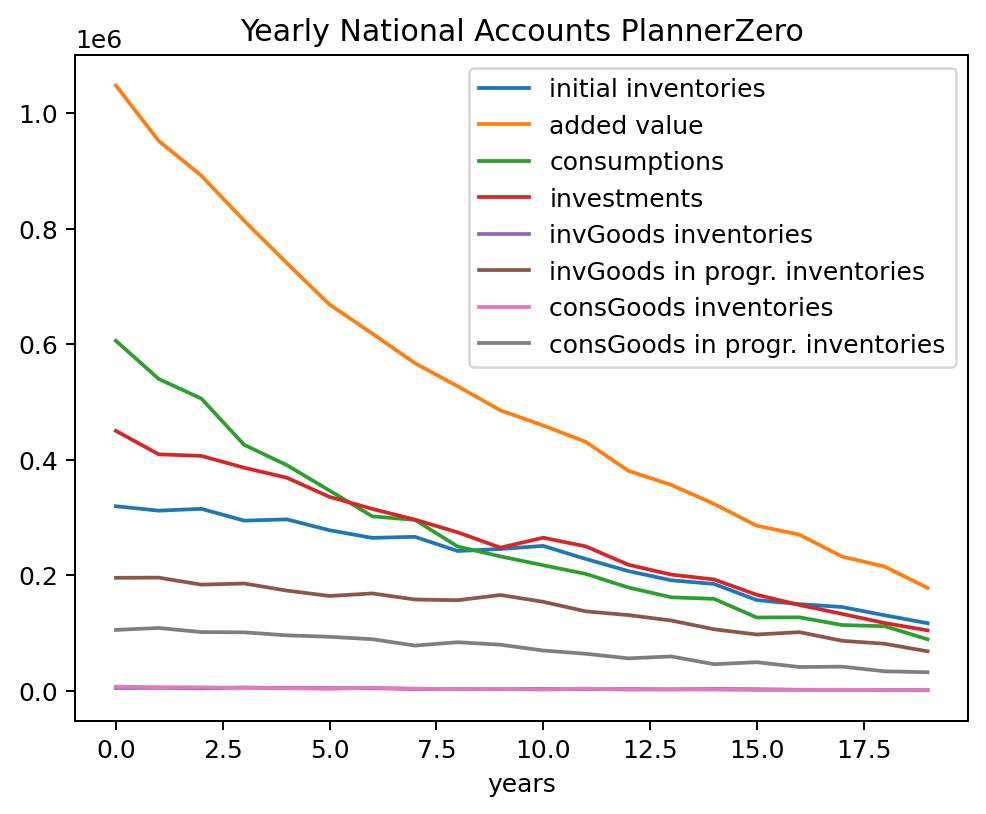}
\end{center}
\caption{National accounts if the planner assigns zero of the investment goods required by the firms}
\label{PlannerZeroNAFig}
\end{figure}

In Fig.~\ref{PlannerZeroNAFig}, we capture the long-run effects of missing investments on production: all the measures describing the economic behavior decline, rather unsurprisingly, towards zero.

%%%%%%%%%%%%%%%%%%%%%%%%%%%%%%%%%%%%%%%%%%%%%%%%%%%%%%%%%%%%%%%%%%%%%%%%%%%%%%%%%%%%%%%
\subsubsection{Planner total}\label{Planner total}

We continue with a second control run, always to verify the model's behavior under stress. The planner accommodates all requests for capital goods from enterprises, regardless of availability; in reality, it should import them.

\begin{figure}[H]
\begin{center}
    \includegraphics[width=0.5\textwidth]{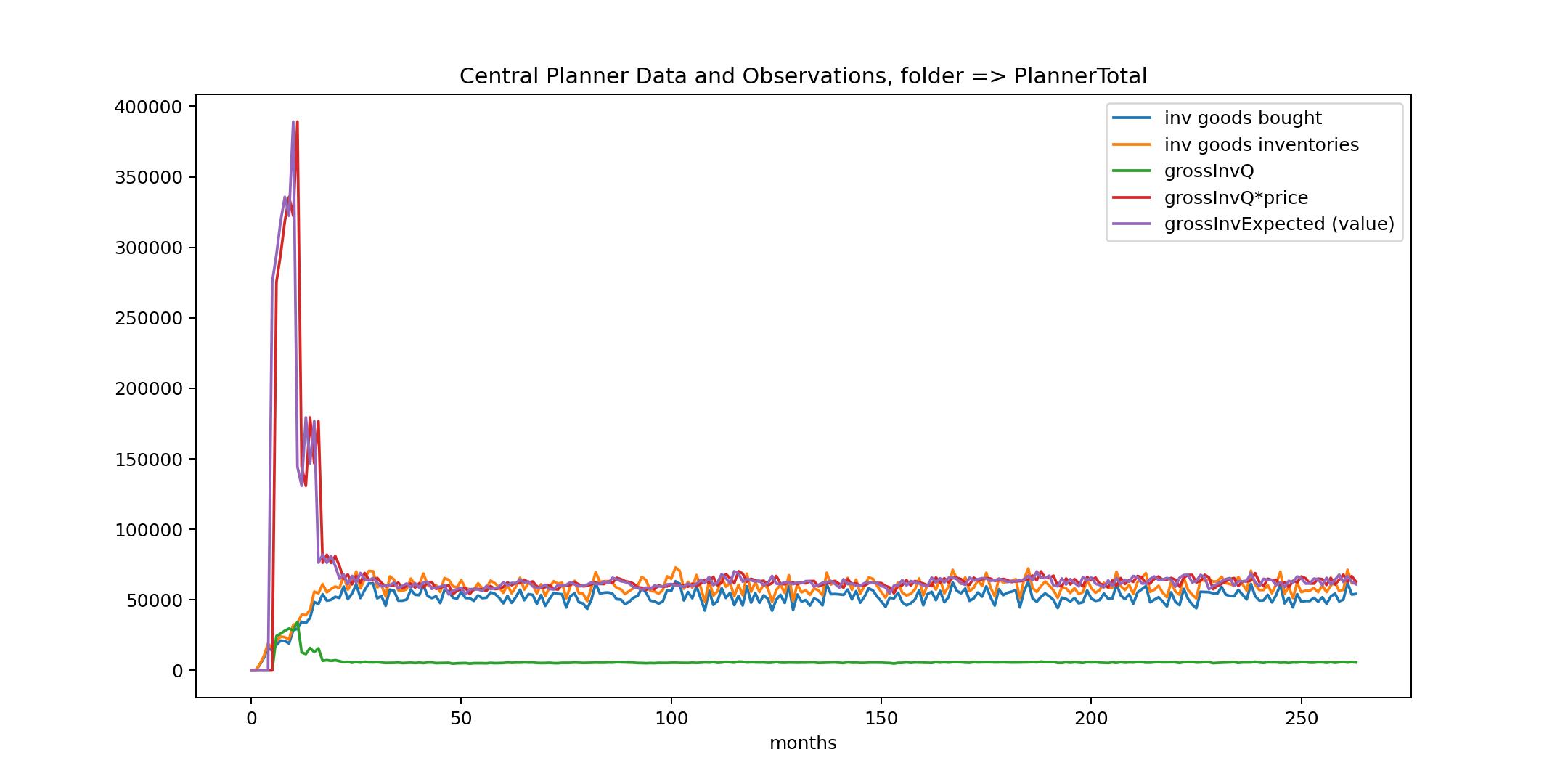}
 \end{center}
\caption{The planner assigns the total quantity of investment goods required by the firms}
\label{PlannerTotalFig}
\end{figure}

\begin{figure}[H]
\begin{center}
\includegraphics[width=0.7\textwidth]{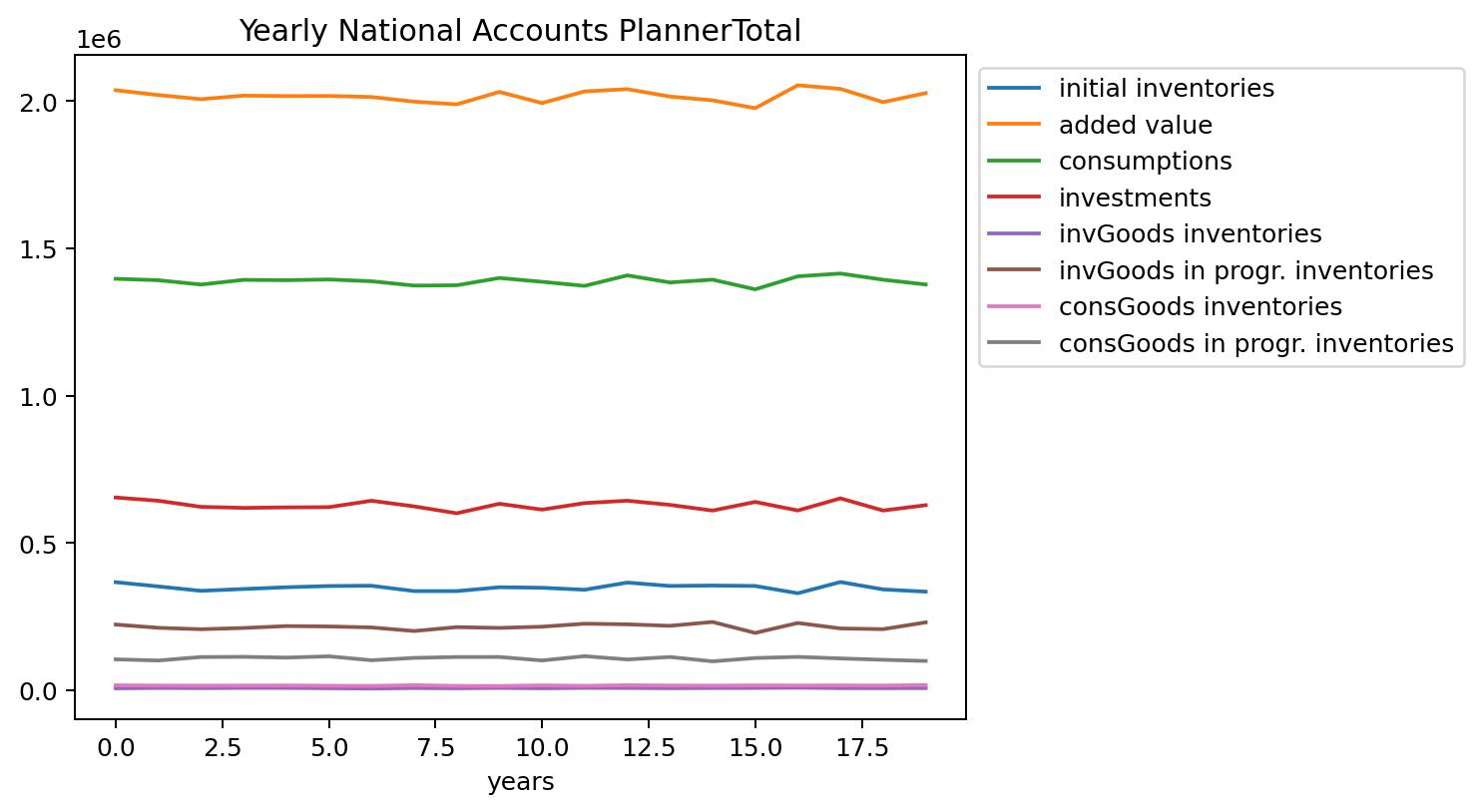}
\end{center}
\caption{National accounts if the planner assigns the total quantity of investment goods required by the firms}
\label{PlannerTotalNAFig}
\end{figure}

In Fig.~\ref{PlannerTotalFig}, the planner behavior plot displays a jump in its early phase as in the previous experiment. The event coincides with a considerable excess of what firms receive (red line) compared with the investment goods that the planner buys (blue line): to overcome this situation, the planner would have to collect imports of capital goods.
After this \emph{warming-up} phase, the model balances expected and assigned investment goods and reaches its stationary state.
The same stationary state emerges from Fig.~\ref{PlannerTotalNAFig}, where all the values are constant over time. It is interesting to notice here that the size of consumption is valuably larger than that of investments: this depends on the structure of the firm population, which presents many more firms producing consumption goods than durable, productive goods.

\emph{\c{C}a va sans dire}, in all the cases, the sum of investments, consumption and inventories variation returns the total added value of the economy.

%%%%%%%%%%%%%%%%%%%%%%%%%%%%%%%%%%%%%%%%%%%%%%%%%%%%%%%%%%%%%%%%%%%%%%%%%%%%%%%%%%%%%%%
\subsubsection{Planner random}\label{Planner random}

With this third control run, always to verify the model's behavior under stress, we introduce a planner reacting to the firms' requests randomly, from 0 to 100\%.

\begin{figure}[H]
\begin{center}
    \includegraphics[width=0.7\textwidth]{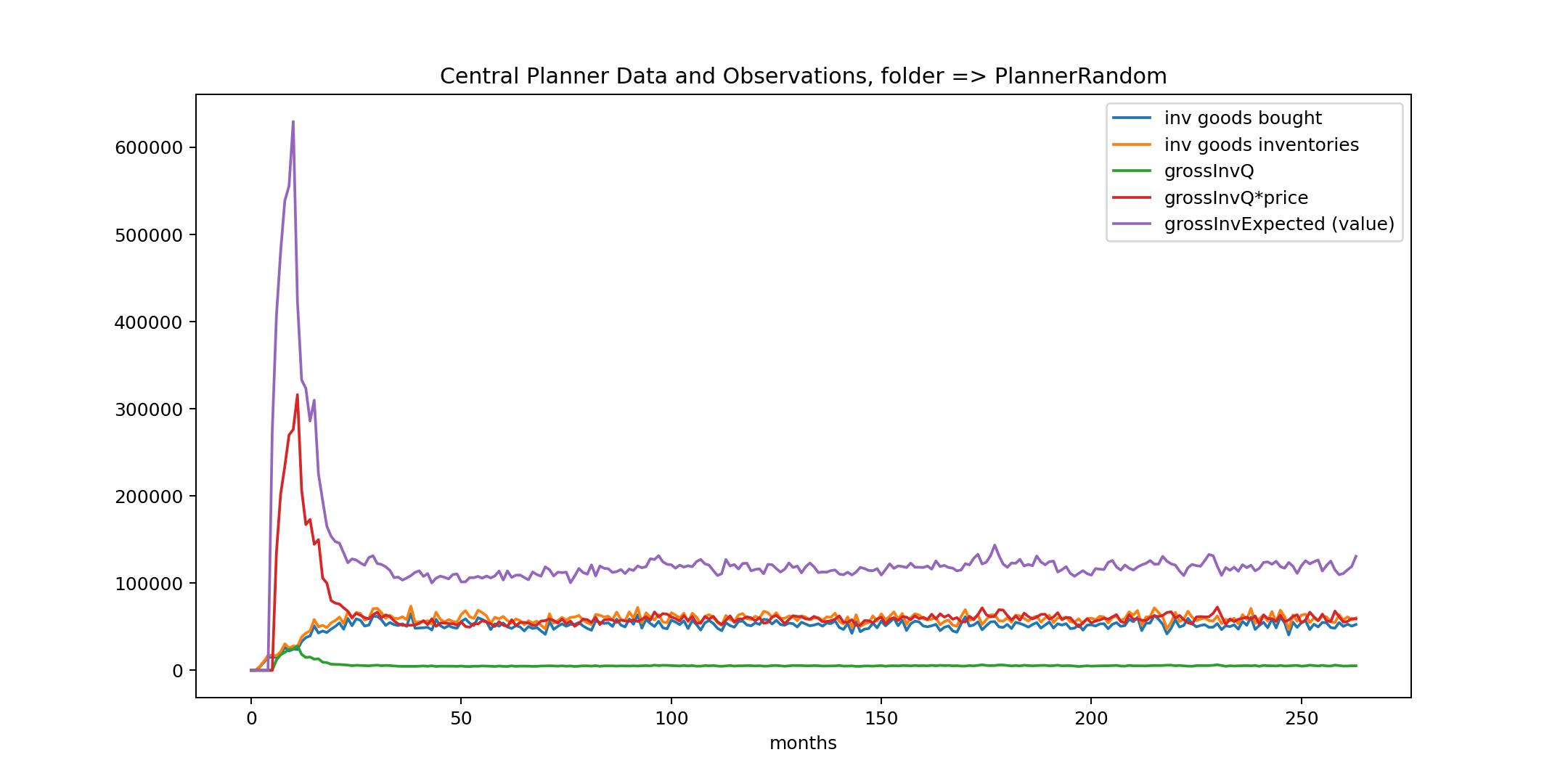}
\end{center}
\caption{The planner assigns a random quantity of investment goods required by the firms}
\label{PlannerRandomFig}
\end{figure}

Fig.~\ref{PlannerRandomFig} shows what happens when the planner distributes only a share of the investment goods that firms require. Moreover, this share is randomly extracted. At the beginning of the simulation, we observe the same initial boost that we saw in the previous cases, which highlights a considerable excess of requests concerning the quantity existing internally in the economy. In this case, we again observe a stable trend in the firms' expectations of investment goods.

\begin{figure}[H]
\begin{center}
\includegraphics[width=0.7\textwidth]{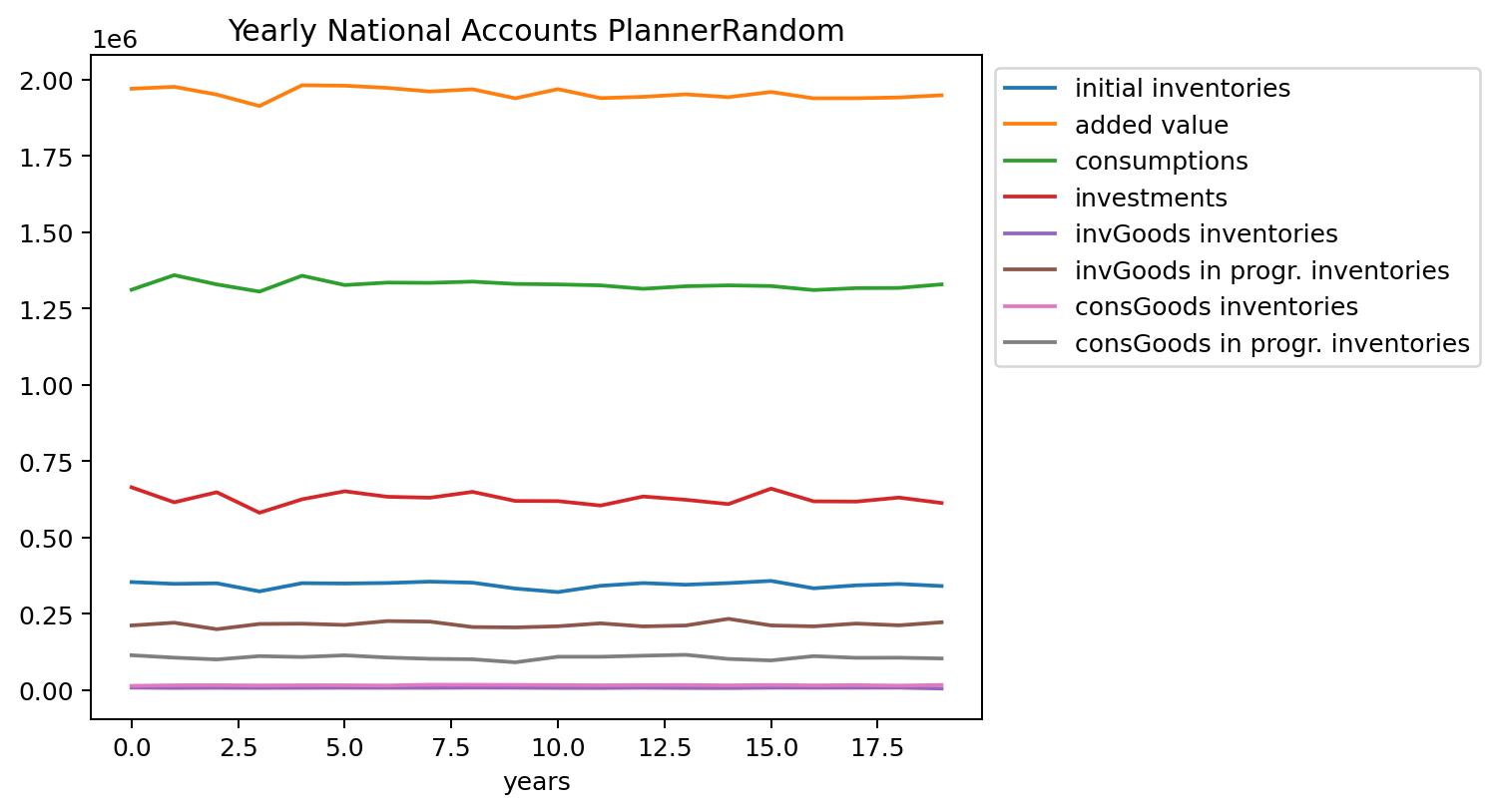}
\end{center}
\caption{National accounts if the planner assigns a random quantity of investment goods required by the firms}
\label{PlannerRandomNAFig}
\end{figure}

It is worth noticing that Fig.~ ef {PlannerRandomNAFig}, reporting the national accounts, presents insignificant differences with the previous case, meaning that the economy is not sensitive to changes in the share of investment goods distributed to the firms. This situation depends on the fact that---being production orders exogenously generated---the productive capacity is still sufficient to fulfill them despite the reduced assignment of investment goods.

%%%%%%%%%%%%%%%%%%%%%%%%%%%%%%%%%%%%%%%%%%%%%%%%%%%%%%%%%%%%%%%%%%%%%%%%%%%%%%%%%%%%%%%
\subsubsection{Planner proportional regular}\label{plannerPropReg1}

We now have the second block of three experiments showing the main results of our model, as we point out in Section \ref{expPlan}. First case: the planner regularly acts without distortions in orders.

\begin{figure}[H]
\begin{center}
    \includegraphics[width=0.7\textwidth]{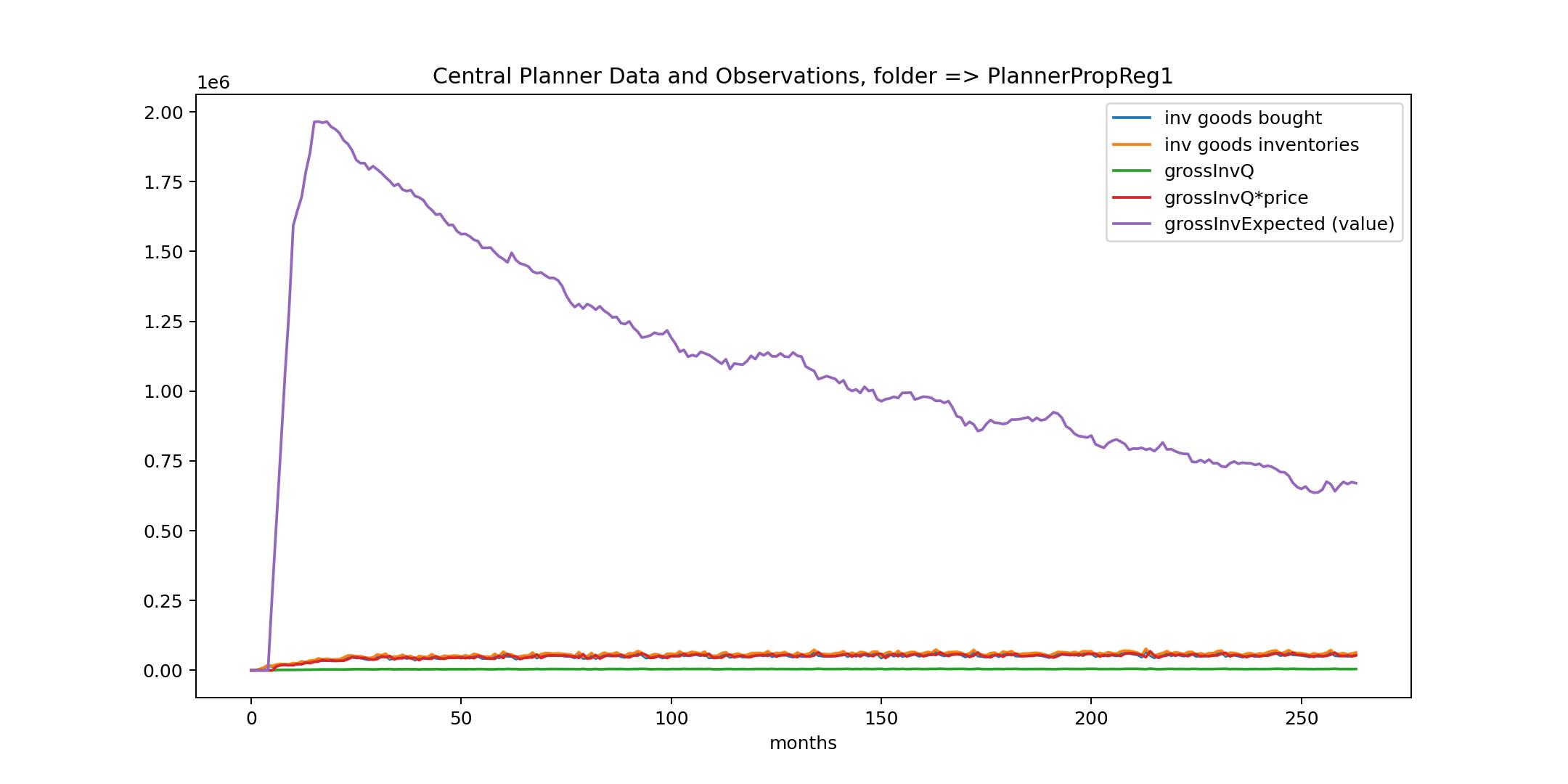}
 \end{center}
\caption{The planner assigns a quantity of investment goods proportional to that required by the firms, there are no distortions in the order generation}
\label{PlannerPropReg1Fig}
\end{figure}

\begin{figure}[H]
\begin{center}
\includegraphics[width=0.7\textwidth]{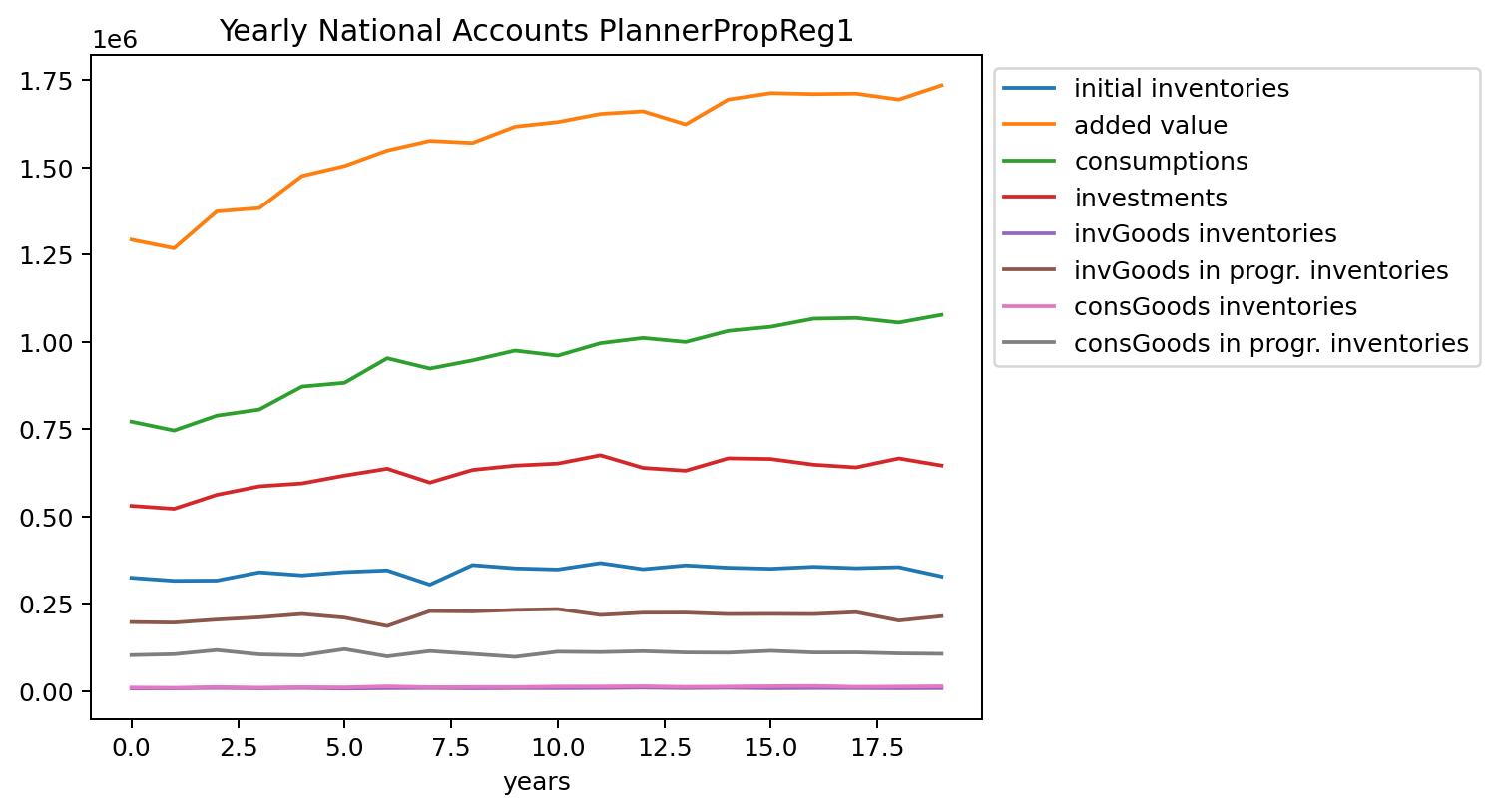}
\end{center}
\caption{National accounts if the planner assigns a quantity of investment goods proportional to that required by the firms, there are no distortions in the order generation}
\label{PlannerPropReg1NAFig}
\end{figure}

In Fig.~\ref{PlannerPropReg1Fig}, we introduce some active behaviors of the central planner as it assigns a quantity of investment goods proportional to that required by the firms under the constraint of the quantity of investment goods bought by the planner previously. This experiment presents a quite realistic scenario as the planner never assigns more than the quantity it buys. This behavior generates a protracted initial excess in the expectations due to the increasing path of the economy.

National accounts in Fig.~\ref{PlannerPropReg1NAFig} experience this significant growth path due to the consistency between firm requests and assigned investments.

%%%%%%%%%%%%%%%%%%%%%%%%%%%%%%%%%%%%%%%%%%%%%%%%%%%%%%%%%%%%%%%%%%%%%%%%%%%%%%%%%%%%%%%
\subsubsection{Planner proportional regular, doubled duration}\label{plannerPropReg2}
\begin{figure}[H]
\begin{center}
    \includegraphics[width=0.7\textwidth]{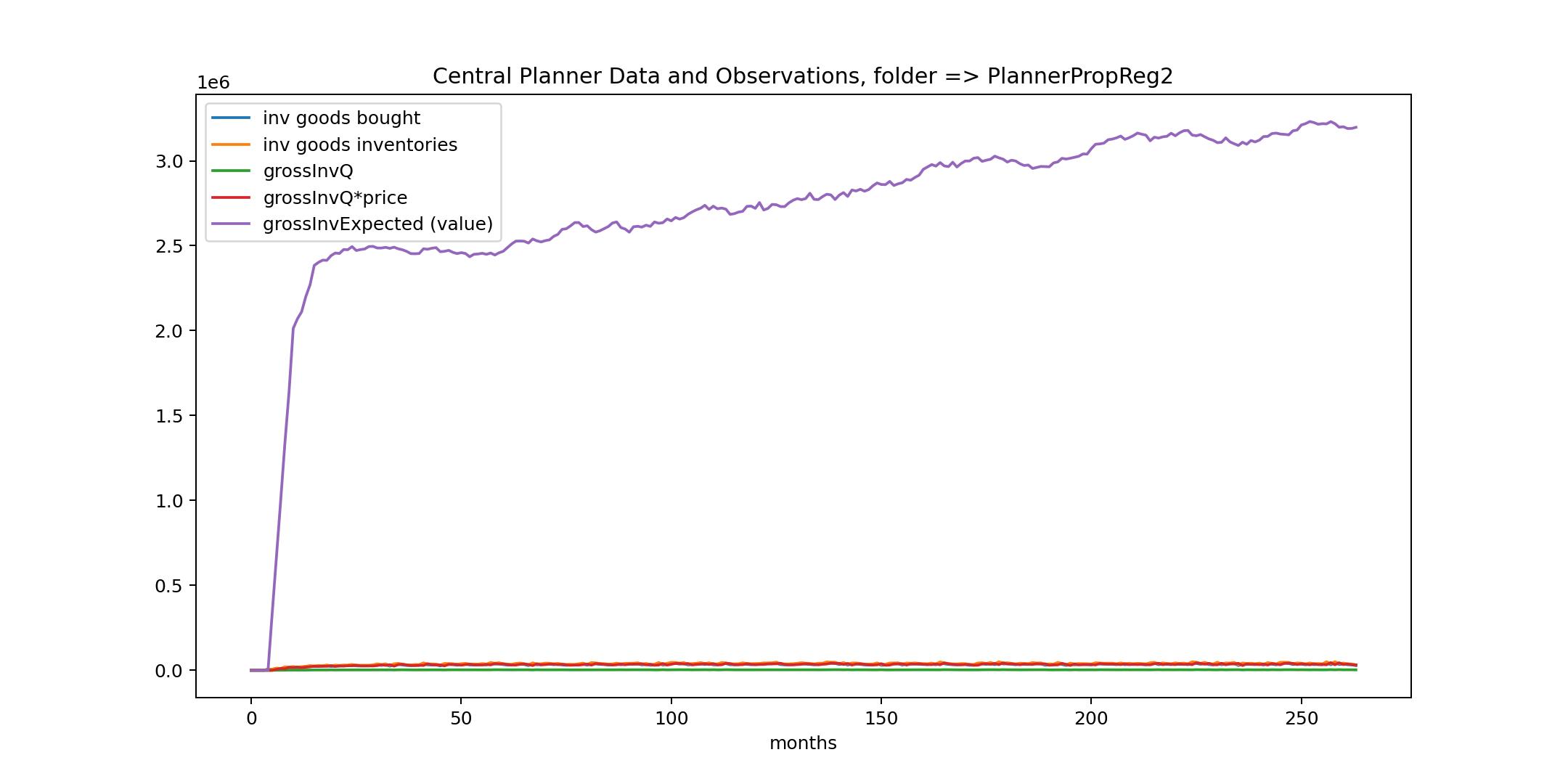}
\end{center}
\caption{The planner assigns a quantity of investment goods proportional to that required by the firms, there are no distortions in the order generation, but the duration is doubled}
\label{PlannerPropReg2Fig}
\end{figure}    

In Fig.~\ref{PlannerPropReg2Fig}, we are analyzing the same case of before, namely the assignment of a quantity of investment goods proportional to that required by the firms under the constraint of the amount of investment goods bought by the planner, but doubling the duration of all the productive processes.
Under this experiment, we observe a significant boost in firm expectations during the first phase. This is a result of sustained production, which is facilitated by the heavy enlargement of production orders due to the doubled duration. This initial boost is followed by a growth trend, albeit at a slower pace, indicating the potential for long-term positive effects.

\begin{figure}[H]
\begin{center}
\includegraphics[width=0.7\textwidth]{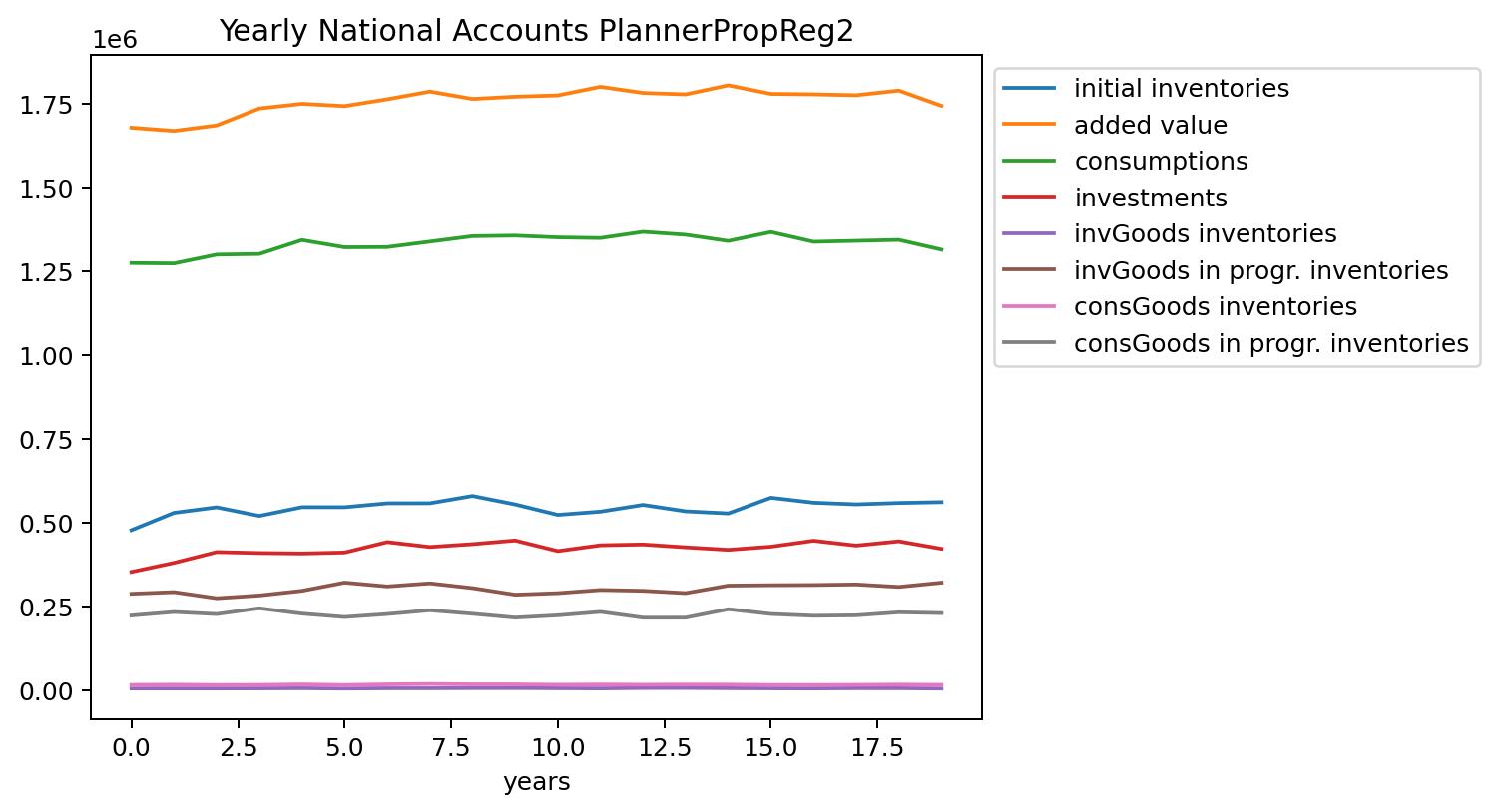}
\end{center}
\caption{National accounts if the planner assigns a quantity of investment goods proportional to that required by the firms, there are no distortions in the order generation, but the duration is doubled}
\label{PlannerPropReg2NAFig}
\end{figure}

Under these assumptions, it is very interesting to analyze the national accounts: in Fig.~\ref{PlannerPropReg2NAFig} they display from the very beginning higher values comparing this scenario with the previous one. The reason for this is the doubling of the duration of the production of those many producers having a small duration: this allows the firms to accept nearly doubled production orders, fulfilling the productive capacity much sooner in the time dynamics.

%%%%%%%%%%%%%%%%%%%%%%%%%%%%%%%%%%%%%%%%%%%%%%%%%%%%%%%%%%%%%%%%%%%%%%%%%%%%%%%%%%%%%%%
\subsubsection{Planner proportional minimising investments}\label{PlannerPropMin1}

We continue with the second block of three experiments, which show the main results of our model. In this second case, the planner introduces distortions in orders, which are now mainly directed at firms producing consumption goods.

\begin{figure}[H]
\begin{center}
    \includegraphics[width=0.7\textwidth]{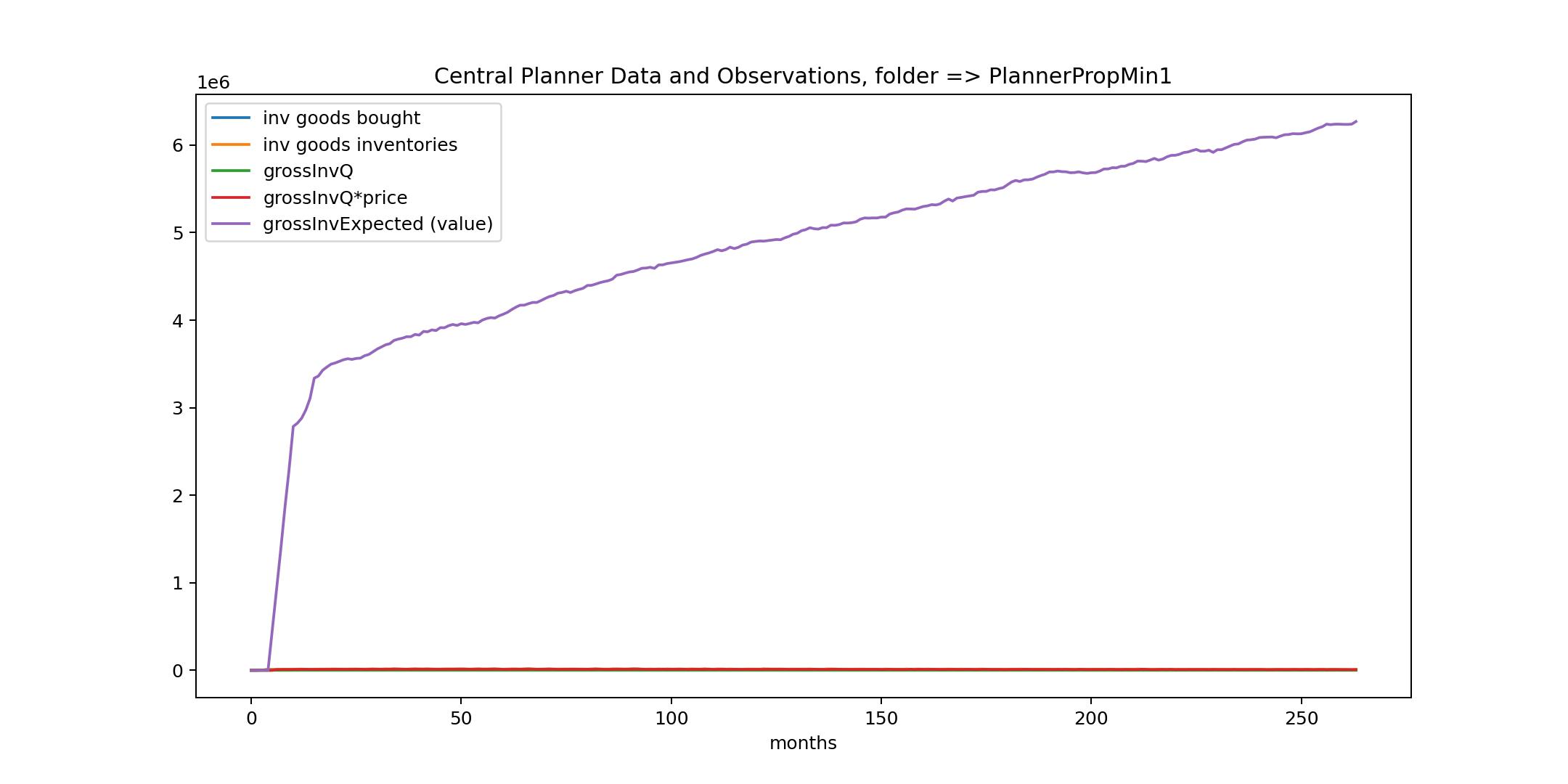}
\end{center}
\caption{The planner assigns a quantity of investment goods proportional to that required by the firms, distorting the market through a \emph{pro-consumption} policy in order generation}
\label{PlannerPropMin1Fig}
\end{figure}    

\begin{figure}[H]
\begin{center}
    \includegraphics[width=0.7\textwidth]{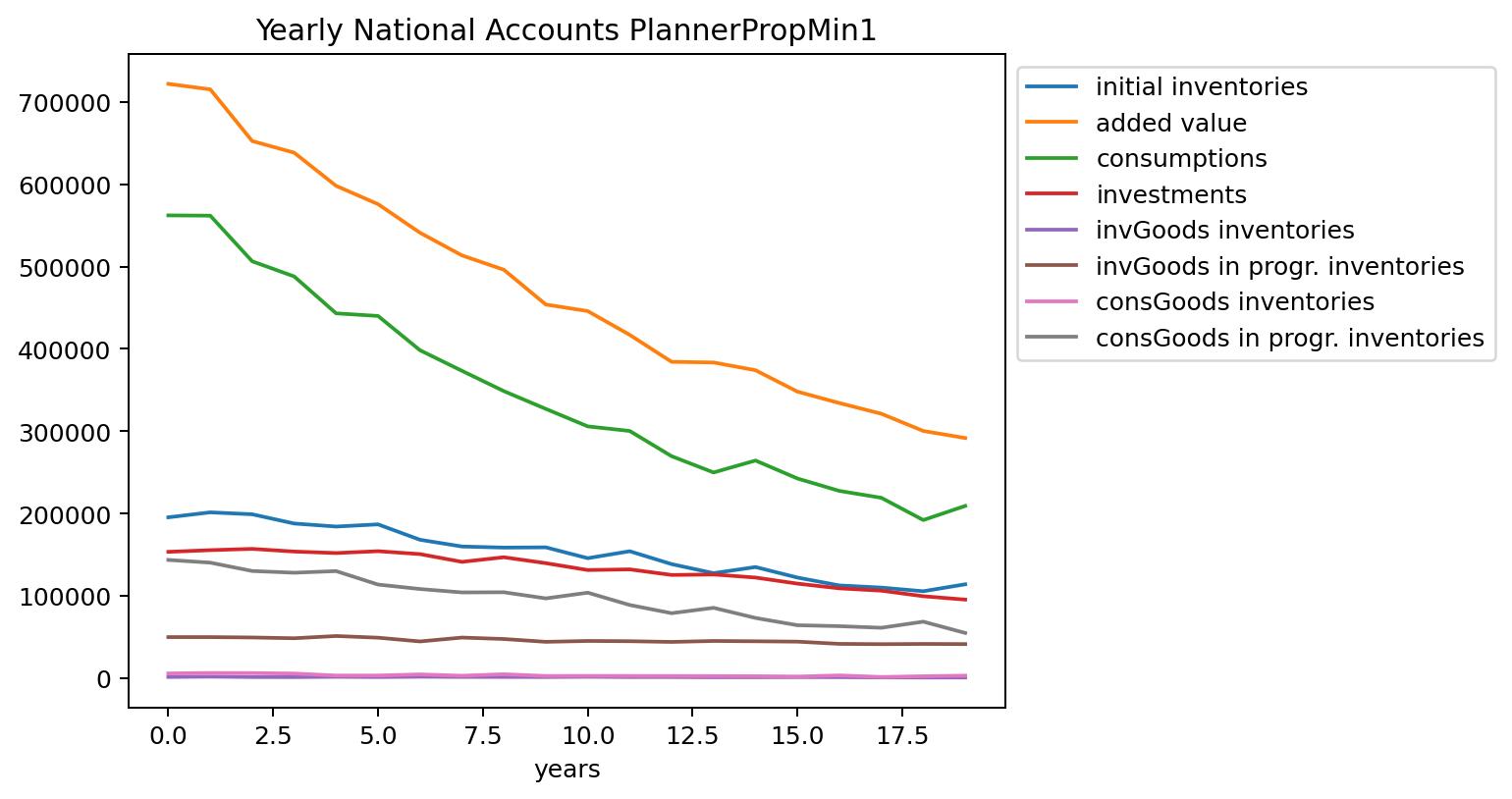}
\end{center}
\caption{National accounts if the planner assigns a quantity of investment goods proportional to that required by the firms, distorting the market through a \emph{pro-consumption} policy in order generation}
\label{PlannerPropMin1NAFig}
\end{figure}

As in the previous pair of cases, in Fig.~\ref{PlannerPropMin1NAFig} the planner assigns a quantity of investment goods proportional to that required by the firms under the constraint of the amount of investment goods bought by the planner in the previous period, but here it applies a \emph{pro-consumption} policy by fostering the order generation for the sectors producing consumption goods.

Observing in Fig.~\ref{PlannerPropMin1Fig} the central planner behavior plot, we have that the weakness of the production of investment goods due to the lack in their order generation causes the central planner not to have the possibility of buying investment goods and then assigning them. This situation yields the firms' infinitely growing unsatisfied demand for investment goods. All of this brings the economy to its collapse, as it emerges in the national accounts plot (please also notice that the scale of values in the plot is dramatically reduced under this hypothesis).

%%%%%%%%%%%%%%%%%%%%%%%%%%%%%%%%%%%%%%%%%%%%%%%%%%%%%%%%%%%%%%%%%%%%%%%%%%%%%%%%%%%%%%%
\subsubsection{Planner proportional minimising investments, doubled duration}\label{PlannerPropMin2}

\begin{figure}[H]
\begin{center}
    \includegraphics[width=0.7\textwidth]{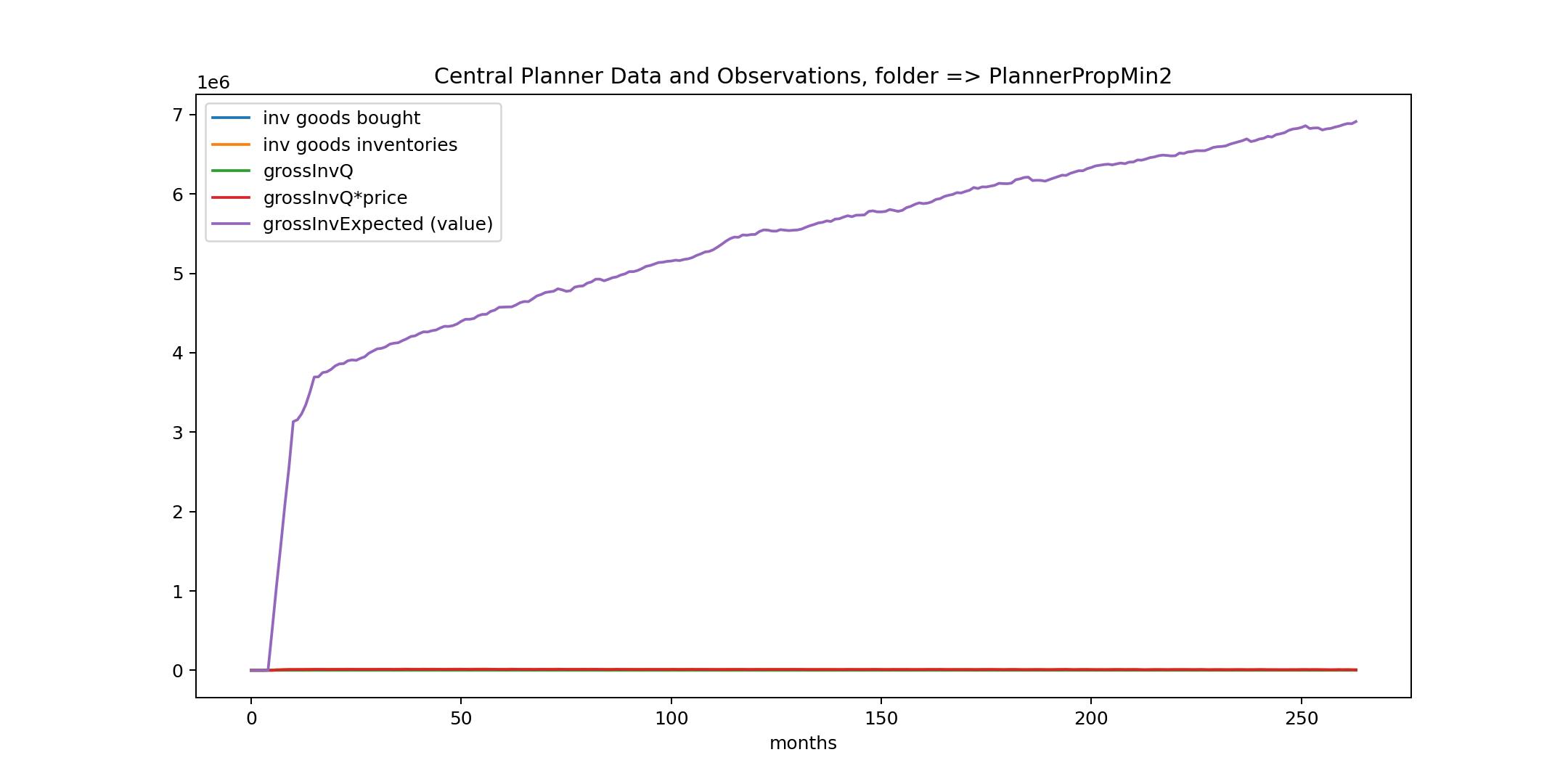}
\end{center}
\caption{The planner assigns a quantity of investment goods proportional to that required by the firms, distorting the market through a \emph{pro-consumption} policy in order generation, but the duration is doubled}
\label{PlannerPropMin2Fig}
\end{figure}

\begin{figure}[H]
\begin{center}
    \includegraphics[width=0.7\textwidth]{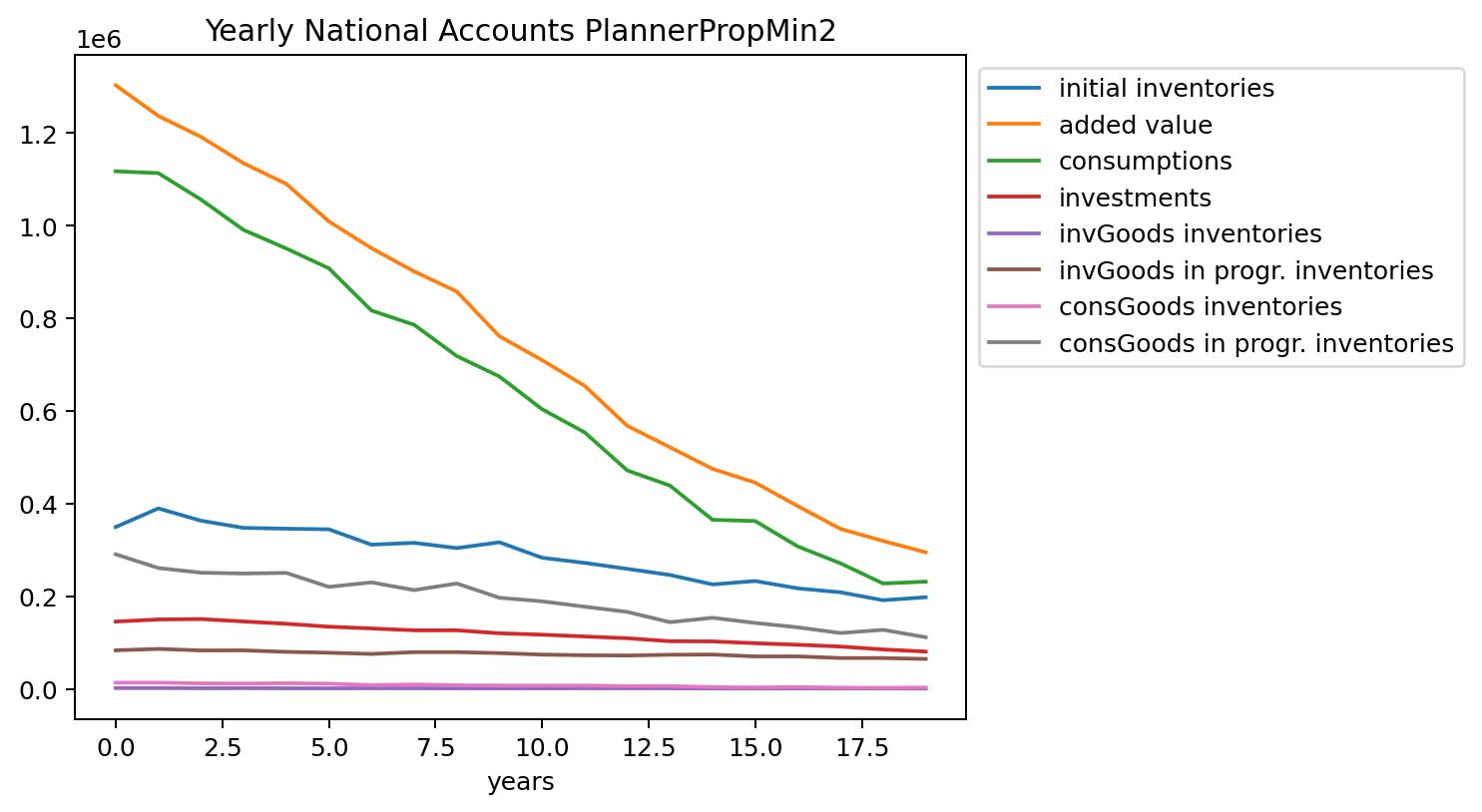}
\end{center}
\caption{National accounts if the planner assigns a quantity of investment goods proportional to that required by the firms, distorting the market through a \emph{pro-consumption} policy in order generation, but the duration is doubled}
\label{PlannerPropMin2NAFig}
\end{figure}

Fig.~\ref{PlannerPropMin2Fig} replicates the scenario of Fig.~\ref{PlannerPropMin1Fig} (\emph{i.e.}, the \emph{pro-consumption} distorting mechanism), but considering a doubled duration of the productive processes. Thus, we observe analogous results, with some slight differences that it is proper to underline. In this case, the economy's starting point is far richer due to the doubled duration (as analyzed in Section \ref{plannerPropReg2}), as we can observe by the dimension of the scale in the national accounts plot. However, the economy's collapse is inexorable under these conditions due to the lack of order generation of productive durable goods, as shown in Fig.~\ref{PlannerPropMin2NAFig}.

%%%%%%%%%%%%%%%%%%%%%%%%%%%%%%%%%%%%%%%%%%%%%%%%%%%%%%%%%%%%%%%%%%%%%%%%%%%%%%%%%%%%%%%
\subsubsection{Planner proportional maximising investments}\label{PlannerPropMax1}

\begin{figure}[H]
\begin{center}
    \includegraphics[width=0.6\textwidth]{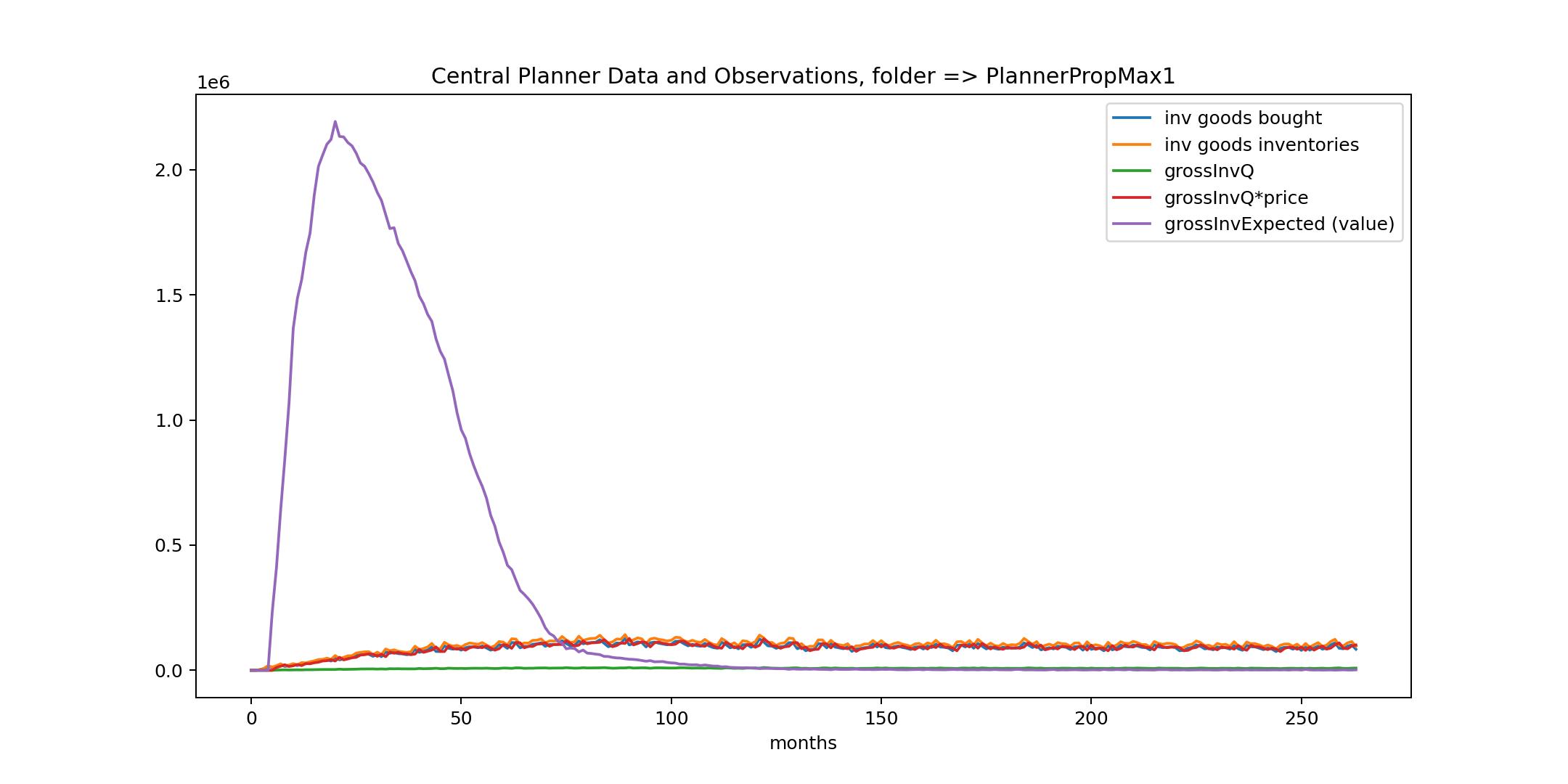}
\end{center}
\caption{The planner assigns a quantity of investment goods proportional to that required by the firms, distorting the market through a \emph{pro-industry} policy in order generation}
\label{PlannerPropMax1Fig}
\end{figure}

\begin{figure}[H]
\begin{center}
    \includegraphics[width=0.7\textwidth]{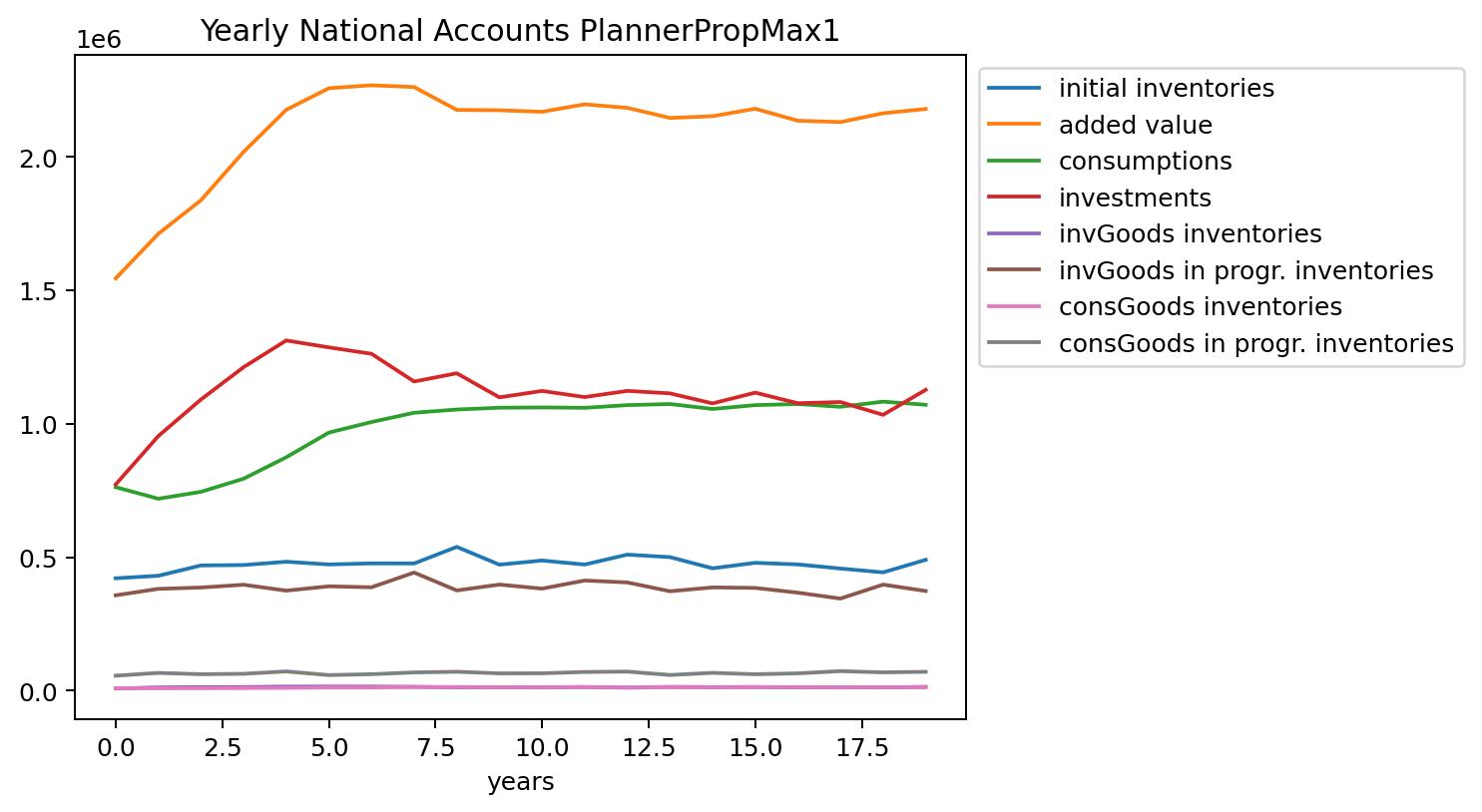}
\end{center}
\caption{National accounts if the planner assigns a quantity of investment goods proportional to that required by the firms, distorting the market through a \emph{pro-industry} policy in order generation}
\label{PlannerPropMax1NAFig}
\end{figure}

We are now in the third block of three experiments, which show the main results of our model. These results are related to a planner that again introduces distortions in orders, now mainly directed at firms producing investment goods.

Fig.~\ref{PlannerPropMax1Fig} proposes a scenario under which the central planner assigns a) a quantity of investment goods that is proportional to that required by the firms, b) constrained by the quantity of investment goods bought by the planner itself, c) distorting the market through a \emph{pro-industry} policy by fostering the order generation for the sectors producing durable capital goods.

After a long initial adapting phase in which the firms' requests for investment goods are extraordinarily higher than the central planner's assignment, the economy stabilizes as it reaches the ceiling of production order generation, whose flow is fully satisfied.

The national accounts Fig.~\ref{PlannerPropMax1NAFig} plot clearly displays this situation, showing higher values for the GDP and an increase in the role of investments, contributing to the GDP in the same measure of consumption. Also, the increase in the inventories is to be noticed.

%%%%%%%%%%%%%%%%%%%%%%%%%%%%%%%%%%%%%%%%%%%%%%%%%%%%%%%%%%%%%%%%%%%%%%%%%%%%%%%%%%%%%%%
\subsubsection{Planner proportional maximising investments, doubled duration}\label{PlannerPropMax2}

\begin{figure}[H]
\begin{center}
    \includegraphics[width=0.7\textwidth]{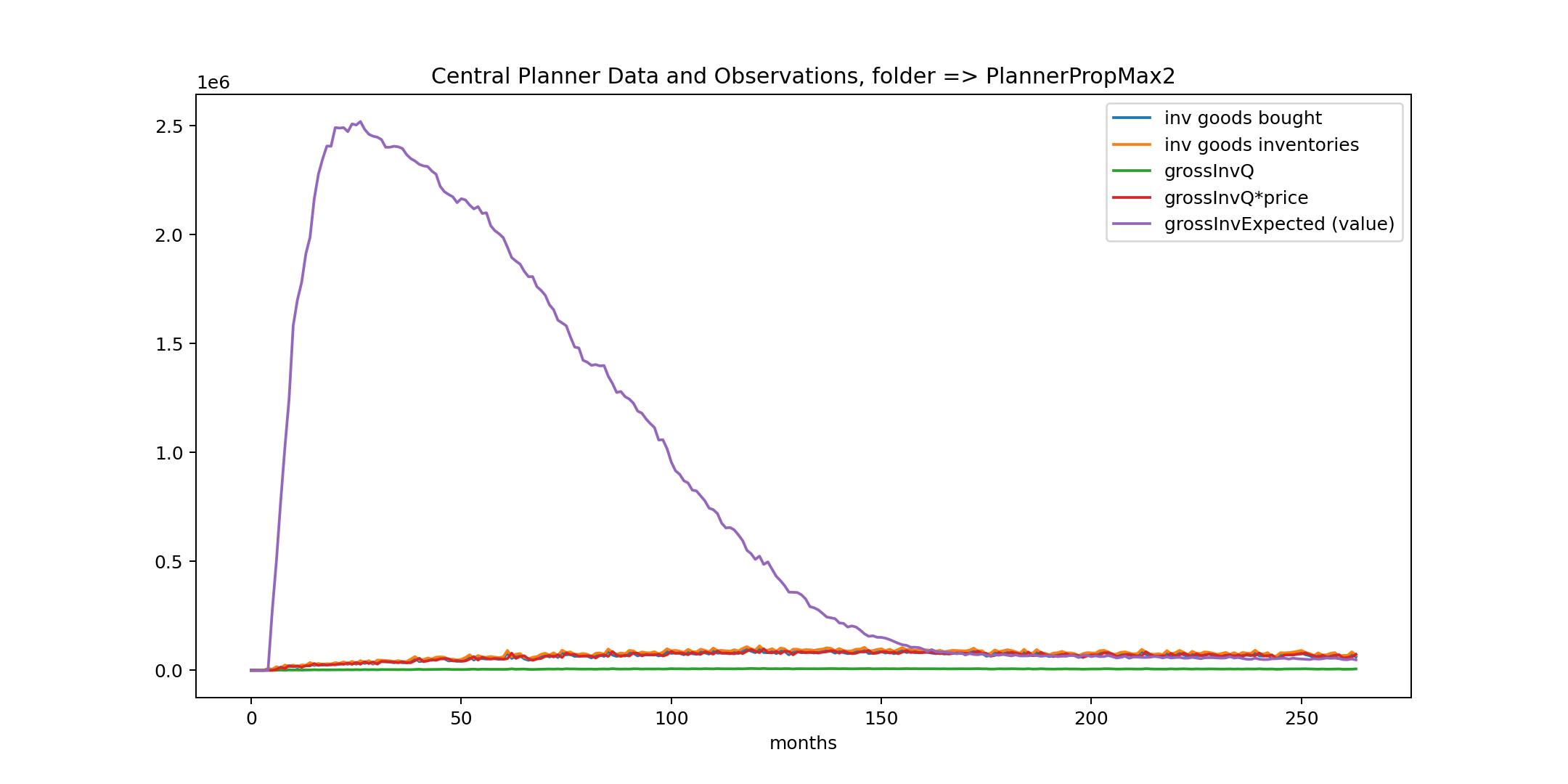}
\end{center}
\caption{The planner assigns a quantity of investment goods proportional to that required by the firms, distorting the market through a \emph{pro-industry} policy in order generation, but the duration is doubled}
\label{PlannerPropMax2Fig}
\end{figure}    

Fig.~\ref{PlannerPropMax2Fig} replicates the scenario of 
Fig.~\ref{PlannerPropMax1Fig} (\emph{i.e.}, the \emph{pro-industry} distorting mechanism), but considering a doubled duration of the productive processes.

\begin{figure}[H]
\begin{center}
    \includegraphics[width=0.7\textwidth]{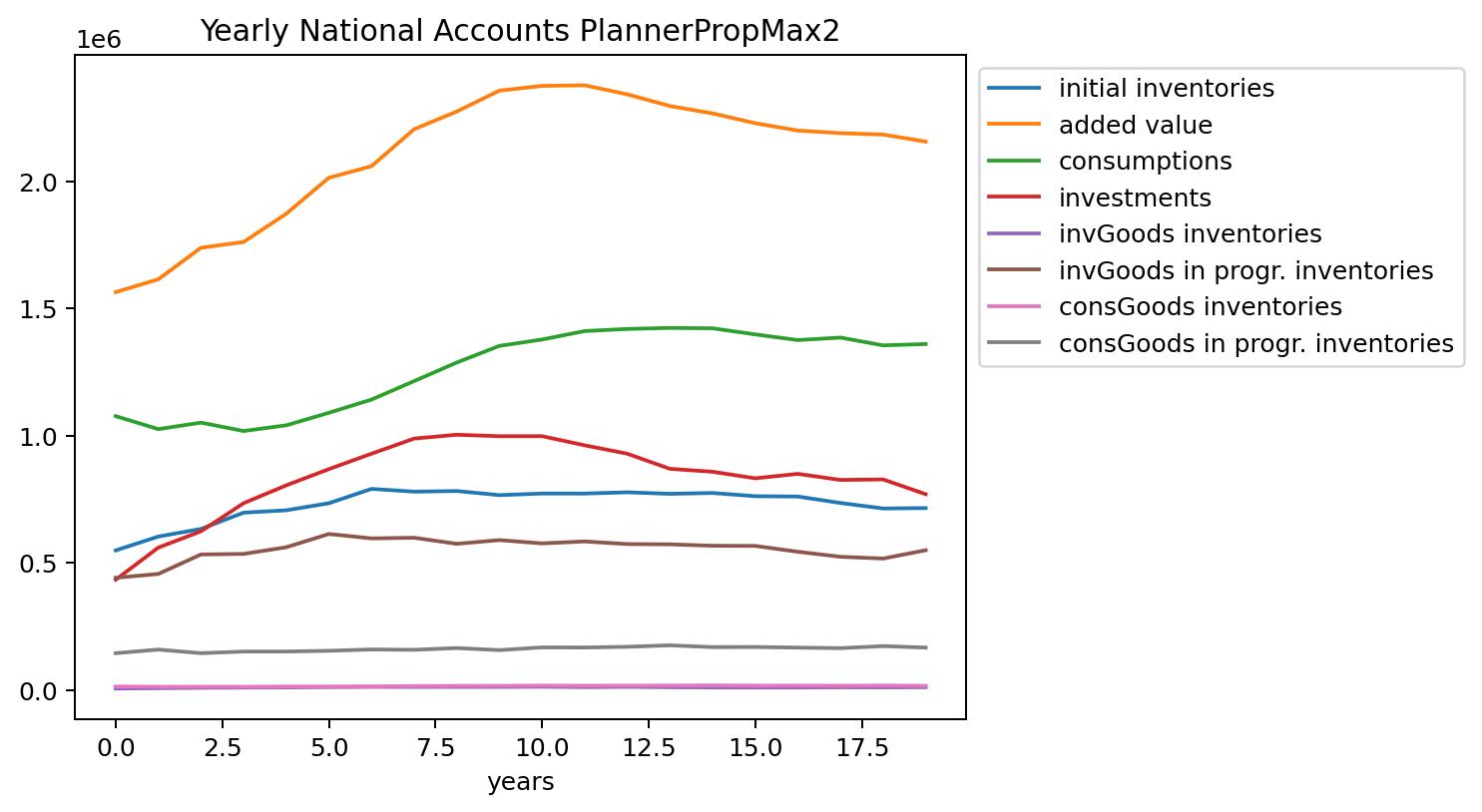}
\end{center}
\caption{National accounts if the planner assigns a quantity of investment goods proportional to that required by the firms, distorting the market through a \emph{pro-industry} policy in order generation, but the duration is doubled}
\label{PlannerPropMax2NAFig}
\end{figure}

In this case, doubling the duration of productive processes results in much more demand for investment goods by the firms with the shortest duration.
However, the increased availability of investment goods for the planner allows a more generous distribution, yielding a boosting phase in the real business cycle. This situation reflects the results in the national accounts plot.
Under this scenario, Fig.~\ref{PlannerPropMax2NAFig}, the proportion between consumption and investments in GDP comes back to be heavily favorable to the former, but with similar global levels, because the increased productive capacity releases the necessary resources to produce consumption goods.

%%%%%%%%%%%%%%%%%%%%%%%%%%%%%%%%%%%%%%%%%%%%%%%%%%%%%%%%%%%%%%%%%%%%%%%%%%%%%%%%%%%%%%%
\subsubsection{Planner proportional maximising investments, failure 0.10}\label{PlannerPropMax1Fail0.1}

\begin{figure}[H]
\begin{center}
    \includegraphics[width=0.7\textwidth]{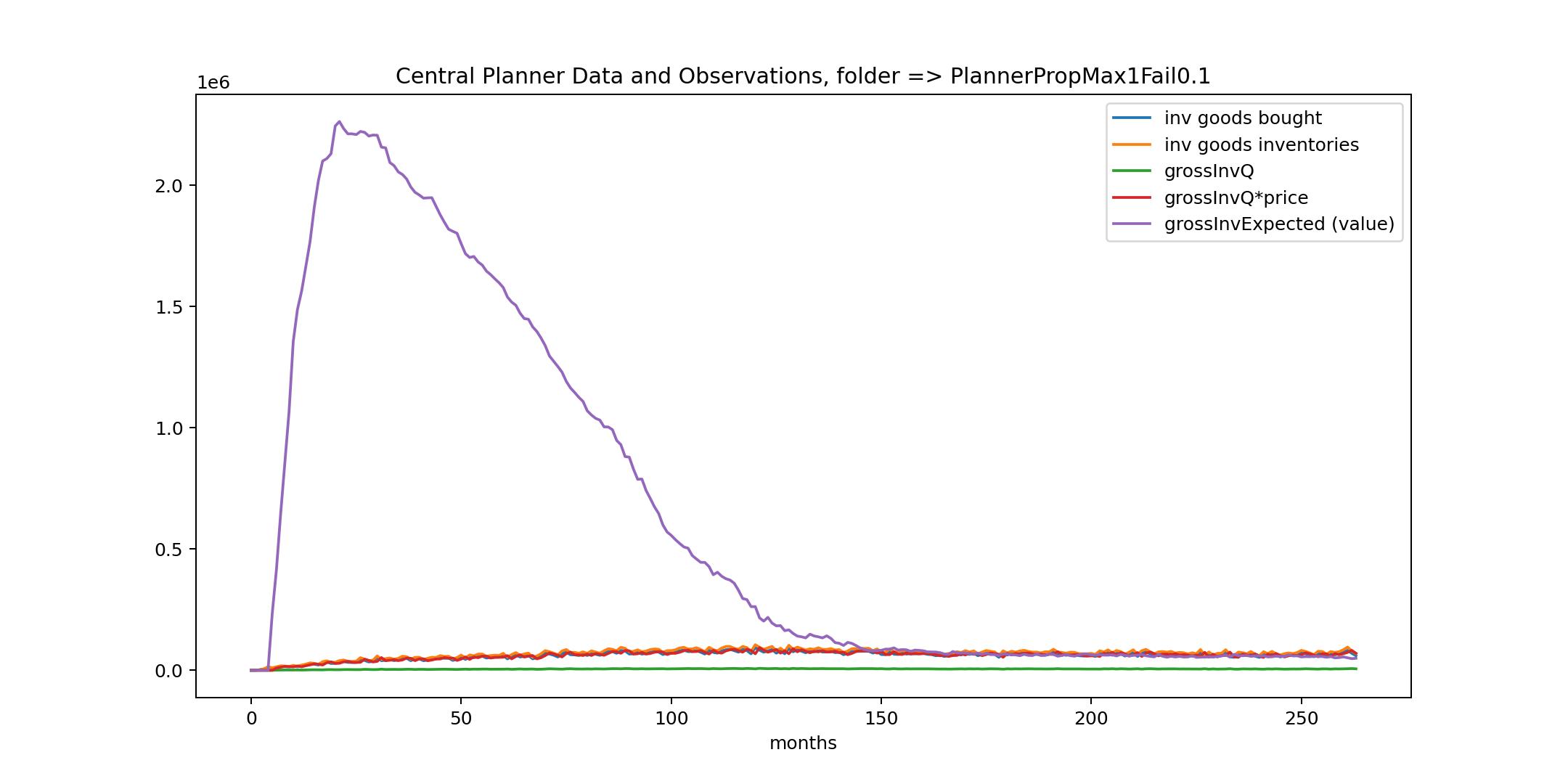}
 \end{center}
\caption{The planner assigns a quantity of investment goods proportional to that required by the firms, distorting the market through a \emph{pro-industry} policy in order generation, but the probability of failure in production increases}
\label{PlannerPropMax1Failure0.1Fig}
\end{figure}

\begin{figure}[H]
\begin{center}
    \includegraphics[width=0.7\textwidth]{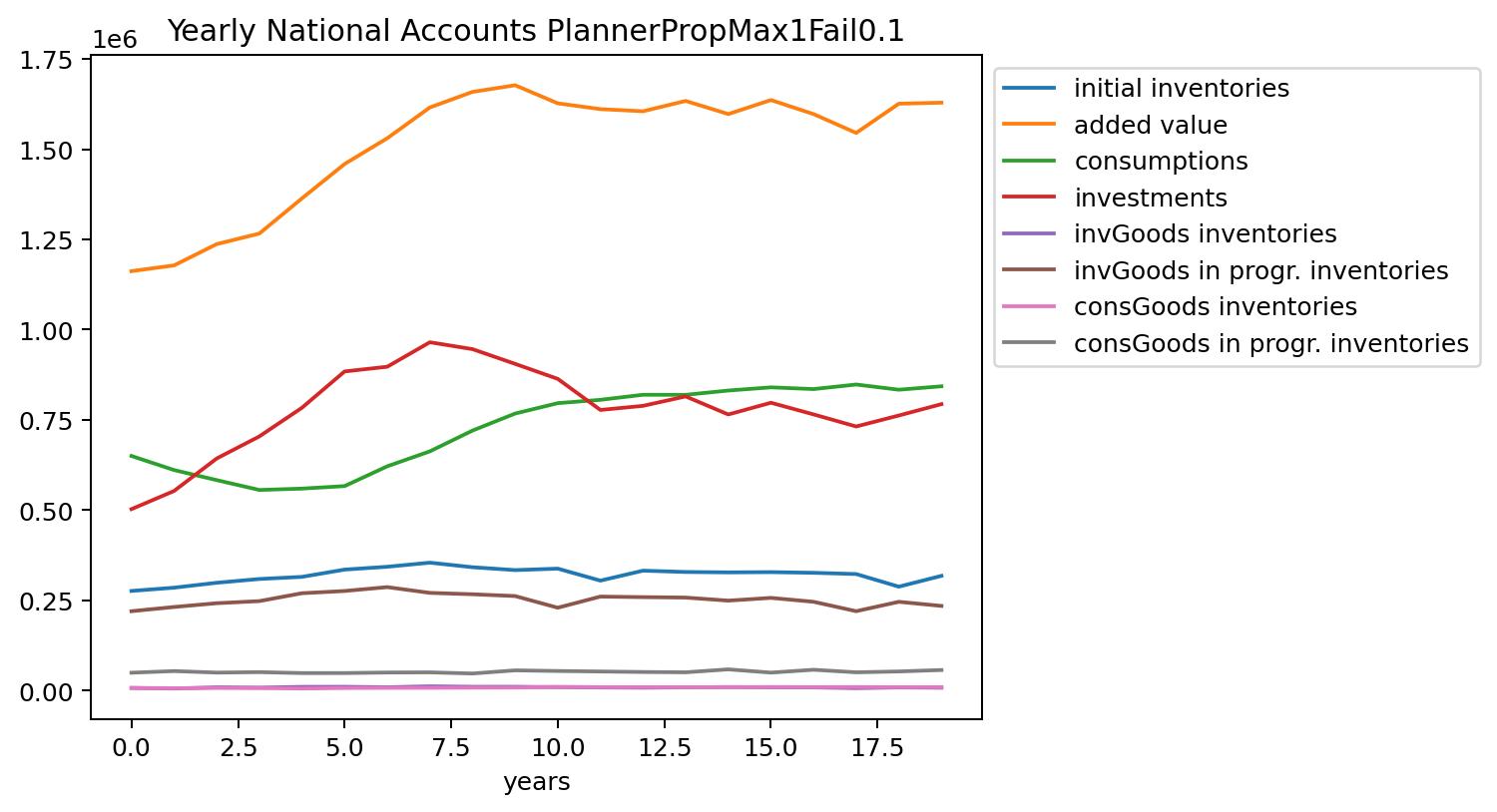}
\end{center}
\caption{National accounts if the planner assigns a quantity of investment goods proportional to that required by the firms, distorting the market through a \emph{pro-industry} policy in order generation, but the probability of failure in production increases}
\label{PlannerPropMax1Failure0.1NAFig}
\end{figure}

We repeat the previous case of investment maximisation, but intervening on another important assumption: here, we increase the parameter that regulates the probability of failures---due to the varying preferences---from 5\% to 10\% to observe changes under this different condition.
The increase in the probability of failure during the productive process 
causes a lack in the availability of investment goods due to the missed accomplishment of their production orders (which are, among other things, the longest ones, generating thus a massive loss in the firm balance sheets). For this reason, the central planner behavior plot displays a significantly higher level of unsatisfied requests for investment goods by firms.

This yields an impressive contraction in the macroeconomic variables reported in the national accounts plot. Here in Fig. \ref{PlannerPropMax1Failure0.1NAFig}, similarly to Fig.~\ref{PlannerPropMax1NAFig}, we observe the pairing between investments and consumption due to the joint effect between the lack of resources to produce both of them and the distorting mechanism pushing up the production of investment goods against that of consumption ones.

%%%%%%%%%%%%%%%%%%%%%%%%%%%%%%%%%%%%%%%%%%%%%%%%%%%%%%%%%%%%%%%%%%%%%%%%%%%%%%%%%%%%%%%
\subsubsection{Planner proportional maximising investments, doubled duration, failure 0.10}\label{PlannerPropMax2Fail0.1}

\begin{figure}[H]
\begin{center}
    \includegraphics[width=0.7\textwidth]{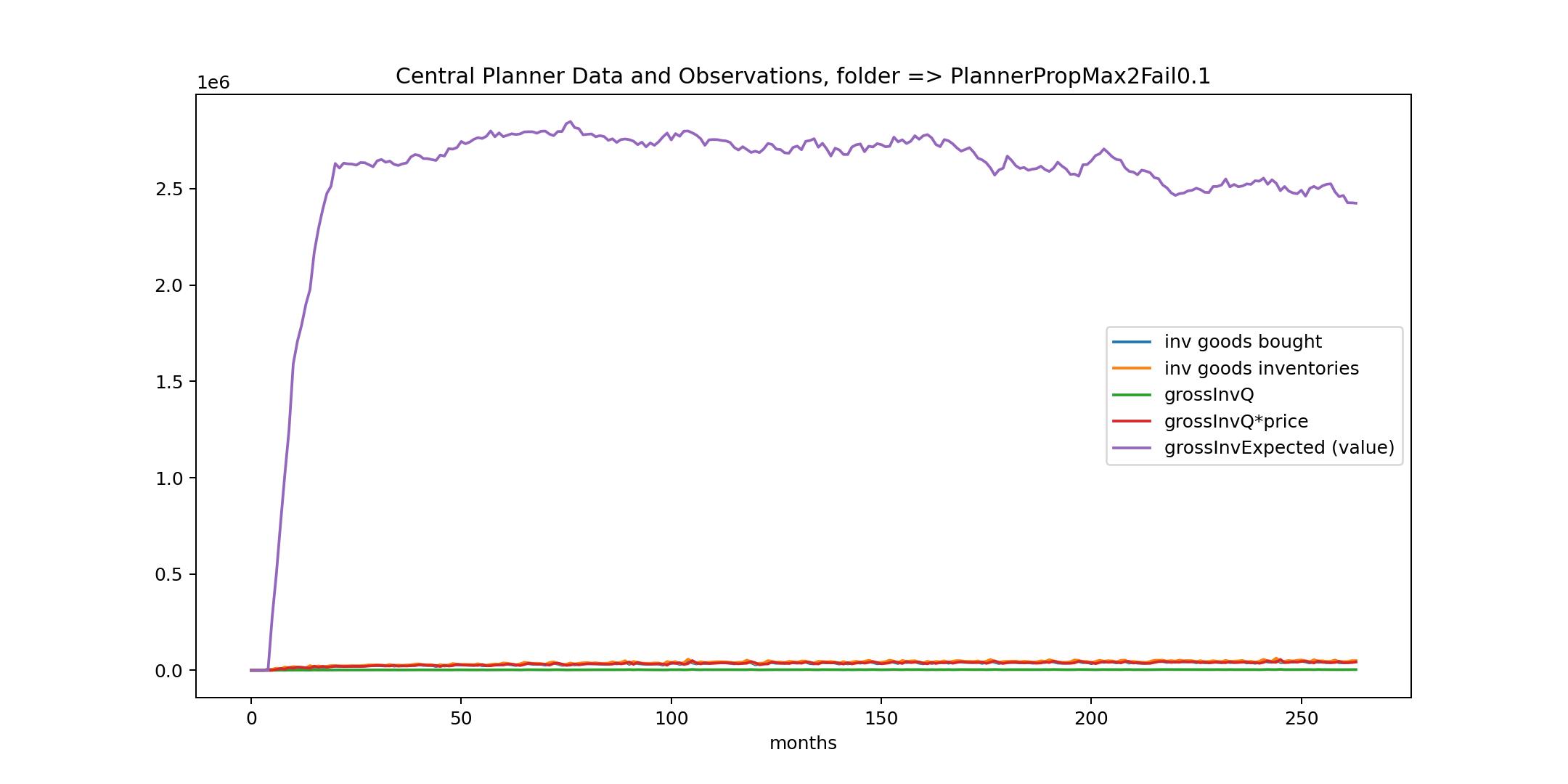}
\end{center}
\caption{The planner assigns a quantity of investment goods proportional to that required by the firms, distorting the market through a \emph{pro-industry} policy in order generation, but the probability of failure in production increases and the duration is doubled}
\label{PlannerPropMax2Failure0.1Fig}
\end{figure}

Fig.~\ref{PlannerPropMax2Failure0.1Fig} replicates the assumptions of Fig.~\ref{PlannerPropMax1Failure0.1Fig} considering a doubled duration of the productive processes. 

Fig.~\ref{PlannerPropMax1Failure0.1Fig} recalls the experiment conducted in Fig.~\ref{PlannerPropMax1Fig} (when the production duration is standard), but intervening on another important assumption: here, we have the doubling of the duration of production processes with a higher risk of failure during the latter, we observe a dramatically high increase in the unsatisfied requests of firms for investment goods, due to this combined effect, that can be observed in the plot of the central planner behavior.

\begin{figure}[H]
\begin{center}
    \includegraphics[width=0.7\textwidth]{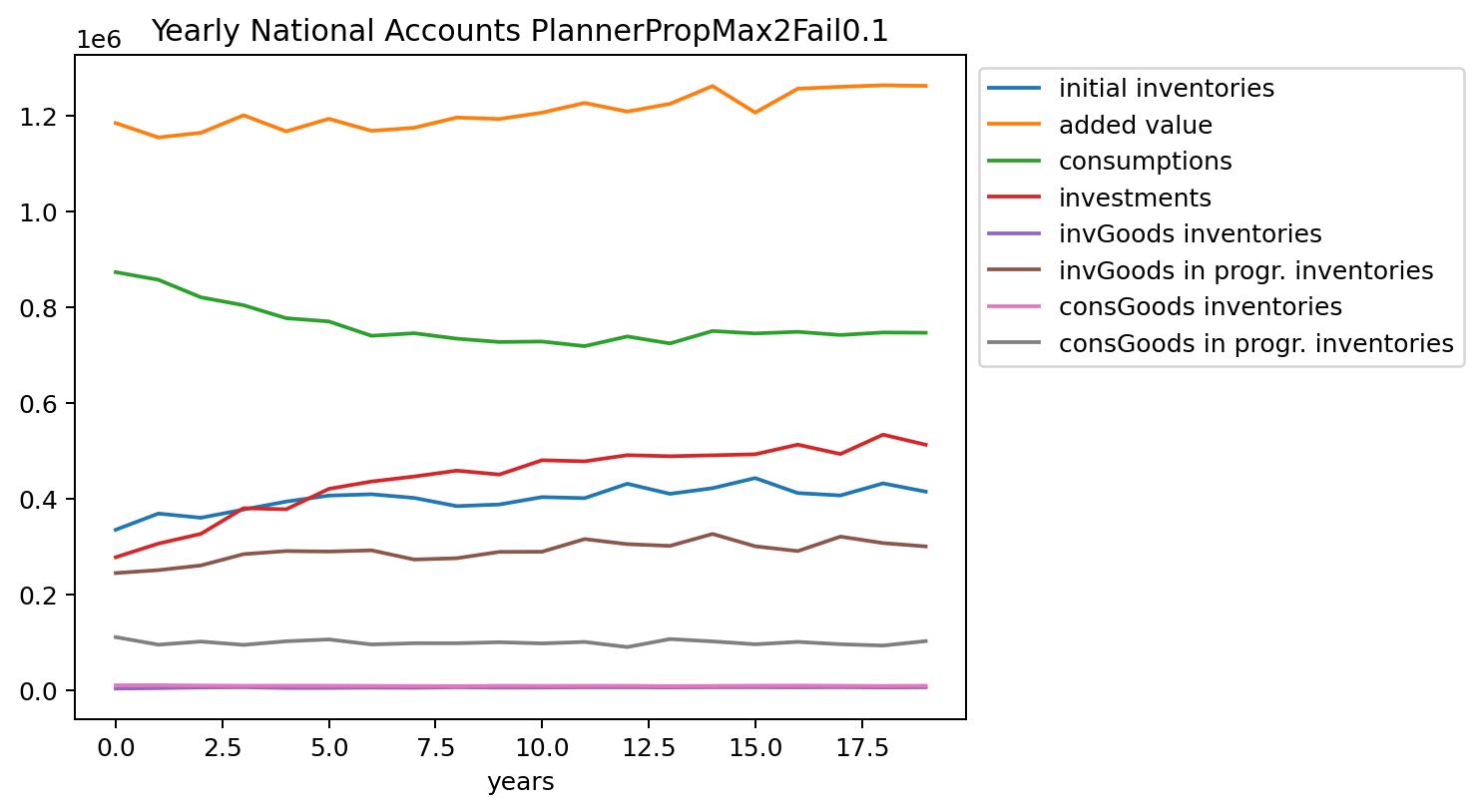}
\end{center}
\caption{National accounts if the planner assigns a quantity of investment goods proportional to that required by the firms, distorting the market through a \emph{pro-industry} policy in order generation, but the probability of failure in production increases and the duration is doubled}
\label{PlannerPropMax2Failure0.1NAFig}
\end{figure}

This phenomenon presents a cascade effect on the general macroeconomic framework, Fig. ~\ref{PlannerPropMax2Failure0.1NAFig}, which suffers from a critical diminishing. However, the increased production flow derived from the doubled duration allows to keep the advantage in the production of consumption goods over investment goods despite the \emph{pro-industry} distorting policy.

%%%%%%%%%%%%%%%%%%%%%%%%%%%%%%%%%%%%%%%%%%%%%%%%%%%%%%%%%%%%%%%%%%%%%%%%%%%%%%%%%%%%%%%
\section{Future developments}

Continuing the journey in the direction of the creation of a whole agent-based grounded macroeconomics, we plan to add: (i) the production of intermediate products with the observation of the related Input-Output Table as a statistical outcome of the model; (ii) the banks and the central bank, the latter with its policies; (iii) the government, with decisions on taxation and public budgeting. Furthermore, a more sophisticated version of the model will include a real market made up by households, who will also interact with the firms on the labor market.

%%%%%%%%%%%%%%%%%%%%%%%%%%%%%%%%%%%%%%%%%%%%%%%%%%%%%%%%%%%%%%%%%%%%%%%%%%%%%%%%%%%%%%%
\bibliography{bibliografiaGenerale}

\begin{thebibliography}{21}
\ProvideTextCommand{\guillemotleft}{OT1}{%
  \leavevmode\raise .27ex\hbox{$\scriptscriptstyle\ll$}}
\ProvideTextCommand{\guillemotright}{OT1}{%
  \leavevmode\raise .27ex\hbox{$\scriptscriptstyle\gg$}}
\newcommand{\enquote}[1]{\guillemotleft#1\guillemotright}
\expandafter\ifx\csname natexlab\endcsname\relax\def\natexlab#1{#1}\fi
\expandafter\ifx\csname url\endcsname\relax
  \def\url#1{{\tt #1}}\fi
\expandafter\ifx\csname urlprefix\endcsname\relax\def\urlprefix{URL }\fi

\bibitem[{Acemoglu and Azar(2020)}]{AcemogluAzart2020}
Acemoglu, D. and Azar, P.~D. (2020).
\newblock {\it Endogenous Production Networks\/}.
\newblock In \enquote{Econometrica}, vol.~88(1), pp. 33-82.
\newline\urlprefix\url{https://onlinelibrary.wiley.com/doi/abs/10.3982/ECTA15899}

\bibitem[{Axtell(2000)}]{axtell2000agents}
Axtell, R. (2000).
\newblock {\it {Why agents? On the varied motivations for agent computing in
  the social sciences}\/}, vol.~17.
\newblock {Center on Social and Economic Dynamics Washington, DC}.

\bibitem[{Barone(2012, Italian edition 1908)}]{10.2307/43828055}
Barone, E. (2012, Italian edition 1908).
\newblock {\it {The Ministry of Production in the Collectivist State}\/}.
\newblock In \enquote{Giornale degli Economisti e Annali di Economia}, vol.
  71(Anno 125)(2/3), pp. 75--112.
\newline\urlprefix\url{http://www.jstor.org/stable/43828055}

\bibitem[{Birner(2001)}]{birner2001cambridge}
Birner, J. (2001).
\newblock {\it {The Cambridge Controversies in Capital Theory; A Study in the
  Logic of Theory Development}\/}.
\newblock In .

\bibitem[{Delli~Gatti {\it et~al.\/}(2011)Delli~Gatti, Desiderio, Gaffeo,
  Cirillo and Gallegati}]{gatti2011macroeconomics}
Delli~Gatti, D., Desiderio, S., Gaffeo, E., Cirillo, P. and Gallegati, M.
  (2011).
\newblock {\it Macroeconomics from the Bottom-up\/}, vol.~1.
\newblock Springer Science \& Business Media.

\bibitem[{Dosi(2023)}]{dosi2023curves}
Dosi, G. (2023).
\newblock {\it {Why is economics the only discipline with so many curves going
  up and down? There is an alternative}\/}.
\newblock LEM Papers Series 2023/02, Laboratory of Economics and Management
  (LEM), Sant'Anna School of Advanced Studies, Pisa, Italy.
\newline\urlprefix\url{https://ideas.repec.org/p/ssa/lemwps/2023-02.html}

\bibitem[{Dosi {\it et~al.\/}(2016)Dosi, Grazzi, Marengo and
  Settepanella}]{dosi2016production}
Dosi, G., Grazzi, M., Marengo, L. and Settepanella, S. (2016).
\newblock {\it Production theory: accounting for firm heterogeneity and
  technical change\/}.
\newblock In \enquote{The Journal of Industrial Economics}, vol.~64(4), pp.
  875--907.

\bibitem[{Fisher(1930)}]{fisher1930}
Fisher, I. (1930).
\newblock {\it The Theory of Interest: As Determined by Impatience to Spend
  Income and Opportunity to Invest It\/}.
\newblock MacMillan, New York.
\newline\urlprefix\url{https://fraser.stlouisfed.org/title/theory-interest-6255}

\bibitem[{Galbraith(2001)}]{galbraith2001distribution}
Galbraith, J.~K. (2001).
\newblock {\it The distribution of income\/}.
\newblock In \enquote{A New Guide to Post Keynesian Economics}, pp. 32--42.

\bibitem[{Georgescu-Roegen(1975)}]{roegen1975model}
Georgescu-Roegen, N. (1975).
\newblock {\it Dynamic models and economic growth\/}.
\newblock In \enquote{World Development}, vol.~3(11), pp. 765-783.
\newline\urlprefix\url{https://www.sciencedirect.com/science/article/pii/0305750X75900790}

\bibitem[{Hayek(1941)}]{hayek2007pure}
Hayek, F.~A. (1941).
\newblock {\it The pure theory of capital\/}.
\newblock Mises Institute printing 2009.

\bibitem[{Hayek({1941/2007})}]{Hayek.2007}
--- ({1941/2007}).
\newblock {\it {The Pure Theory of Capital; first published by Macmillan \& Co.
  in 1941}\/}, vol.~12 of {\it {The Collected Works of F.~A. Hayek}\/}.
\newblock University of Chicago Press.

\bibitem[{Hayek(1981,2012)}]{hayek19812012}
--- (1981,2012).
\newblock {\it {``The Flow of Goods and Services,'' reprinted in Business
  Cycles: Part II, ed. Hansjoerg Klausinger, vol. 8 of The Collected Works of
  FA Hayek}\/}.

\bibitem[{Iannino {\it et~al.\/}(2020)Iannino, Mocci, Vannocci, Colla, Caputo
  and Ferraris}]{app10124343}
Iannino, V., Mocci, C., Vannocci, M., Colla, V., Caputo, A. and Ferraris, F.
  (2020).
\newblock {\it An Event-Driven Agent-Based Simulation Model for Industrial
  Processes\/}.
\newblock In \enquote{Applied Sciences}, vol.~10(12).
\newline\urlprefix\url{https://www.mdpi.com/2076-3417/10/12/4343}

\bibitem[{Lau {\it et~al.\/}(2006)Lau, Huang, Mak and Liang}]{supplyChains2006}
Lau, J., Huang, G., Mak, K. and Liang, L. (2006).
\newblock {\it Agent-based modeling of supply chains for distributed
  scheduling\/}.
\newblock In \enquote{IEEE Transactions on Systems, Man, and Cybernetics - Part
  A: Systems and Humans}, vol.~36(5), pp. 847-861.

\bibitem[{Mazzoli {\it et~al.\/}(2019)Mazzoli, Morini and
  Terna}]{mazzoli_morini_terna_2019}
Mazzoli, M., Morini, M. and Terna, P. (2019).
\newblock {\it Rethinking Macroeconomics with Endogenous Market Structure\/}.
\newblock Cambridge University Press.

\bibitem[{Mazzucato and Perez(2015)}]{Mazzuccato2015}
Mazzucato, M. and Perez, C. (2015).
\newblock {\it {229Innovation as Growth Policy: The Challenge for Europe}\/}.
\newblock In {\it {The Triple Challenge for Europe: Economic Development,
  Climate Change, and Governance}\/}. Oxford University Press.
\newline\urlprefix\url{https://doi.org/10.1093/acprof:oso/9780198747413.003.0009}

\bibitem[{Nikiforos and Zezza(2017)}]{JOES:JOES12221}
Nikiforos, M. and Zezza, G. (2017).
\newblock {\it STOCK-FLOW CONSISTENT MACROECONOMIC MODELS: A SURVEY\/}.
\newblock In \enquote{Journal of Economic Surveys}, vol.~31(5), pp. 1204--1239.
\newline\urlprefix\url{http://dx.doi.org/10.1111/joes.12221}

\bibitem[{Pangallo(2020)}]{pangallo2020synchronization}
Pangallo, M. (2020).
\newblock {\it Synchronization of endogenous business cycles\/}.
\newline\urlprefix\url{https://arxiv.org/abs/2002.06555?s=03}

\bibitem[{Parv {\it et~al.\/}(2019)Parv, Deaky, Nasulea and Oancea}]{parv2019}
Parv, L., Deaky, B., Nasulea, M.~D. and Oancea, G. (2019).
\newblock {\it Agent-Based Simulation of Value Flow in an Industrial Production
  Process\/}.
\newblock In \enquote{Processes}, vol.~7(2).
\newline\urlprefix\url{https://www.mdpi.com/2227-9717/7/2/82}

\bibitem[{Tsiang(1949)}]{tsiang1949}
Tsiang, S.-C. (1949).
\newblock {\it Rehabilitation of time dimension of investment in macrodynamic
  analysis\/}.
\newblock In \enquote{Economica}, vol.~16(63), pp. 204--217.
\newline\urlprefix\url{https://www.jstor.org/stable/pdf/2549678.pdf}

\end{thebibliography}

\bibliographystyle{plainnatmm}

%%%%%%%%%%%%%%%%%%%%%%%%%%%%%%%%%%%%%%%%%%%%%%%%%%%%%%%%%%%%%%%%%%%%%%%%%%%%%%%%%%%%%%%
\begin{appendices}

%%%%%%%%%%%%%%%%%%%%%%%%%%%%%%%%%%%%%%%%%%%%%%%%%%%%%%%%%%%%%%%%%%%%%%%%%%%%%%%%%%%%%%%
\section{The structure and dynamic of labor and capital}\label{appLK}

The basis: each firm has a certain amount of labor, $L$, and productive capital, $K$, to be employed according to a ratio specific to that firm, i.e., its productive recipe $\frac{K}{L}$.

Considering the \emph{balancesheet} of that firm, we have the scheme at Table \ref{bs} 

\vspace{1.5cm}
\begin{table}[htbp]
\centering
 \begin{tabular}{ p{2.5cm} | p{2.5cm} } 
 assets & liabilities \\
 \hline
 \\
 productive\\capital $K$ & \ldots  \\ 
 \\
 \end{tabular}
\caption{The productive capital as an element of the balancesheet}
\label{bs}
\end{table}
\vspace{1.5cm}

We set the initial amount of $K$ at the time of firm creation.\footnote{As set in Appendix \ref{appAdapt}, $K^q_t$ is the current capital in quantity, with $K^q_0  = K_0/p^K_0$ where $p^K_0$ is the initial price of capital goods; the initial price is set internally by the program, estimating a value consistent with the costs of the firms producing goods for investments, considering also their markup. Temporary, the investment goods are supposed to be produced by a unique sector. The updated mean price is indirectly calculated with $p^K_t=K_t/K^q_t$ as $\Delta K^q$ is determining $\Delta K$ using the updated capital goods price.}
For each $m$ time unit (months, weeks,\dots), the firm verifies if the quantity of workers $L$ is consistent with the last $m$ received orders, considering the work productivity. Consistently, the firm hires new workers or fires some of the present ones, adapting $L$. In the same direction, it would be reasonable to adapt $K$, considering the firm recipe.

We can increase the productive capital of any amount by purchasing additional assets. Instead, the way to decrease the productive assets is (1) by selling them if a transaction is possible or (2) by waiting for substitutions due to physical or economic obsolescence and not realizing them. We follow the option (2).

We do not follow the accounting practice of the amortization, considering directly the substitutions, as in national accounts.

In Table \ref{is} we have the \emph{income statement} of the firm.

\begin{table}[H]
\centering
 \begin{tabular}{ p{2.5cm} | p{2.5cm} } 
 costs & revenues \\
 \hline
 \\
compensation of productive capital & \ldots~~considering markup (optional) in inventory evaluation\\  
wages & \\ 
 &  \\  
substitutions\tablefootnote{Substitutions are related to the current $K_t$ value of the capital; if the capital increment $Delta$ exceeds the substitutions with {$K_{t+1} > K_t$}, the increasing amount of the substitutions is not anyway an element of the current \emph{income statement} of the firm.} & \\
 & \\ 
profits? & \\  
 \end{tabular}
\caption{The cost of the productive capital and the substitutions as elements of the \emph{income statement}}
\label{is}
\end{table}

The \emph{compensation of productive capital} is the sum of interests and rents. We calculate it as $K~r$ with $r$ as \emph{cost of capital}.

Considering the assets' useful life $u$ expressed in years and a \emph{time unit} as a fraction $n$ of one year, the substitution costs (which replace the amortizations) per time unit $t$ is $\frac{K_t}{u~n}$.

\vspace{1cm}
\textbf{A note on actual quantities}

With $r$ as capital compensation and $w$ as labor one, we analyze their actual values as follows.

We set the \emph{wage} for each worker to have value $1$ per time unit.

$r$ is set as an annual rate: e.g., $0.10$. It has to be expressed in terms of time unit, with a scheme of simple interest, the $r$ value expressed for a single period, as a fraction $n$ of one year, is $\frac{r}{n}$.

To set a proportion, e.g., of $\frac{1}{2}$ and $\frac{1}{2}$ for the global compensations of labor and productive capital, we need a recipe $\frac{K}{L}$ with, in a time unit:

\vspace{0.2cm}
$\frac{Kr}{n}=Lw$

\vspace{0.2cm}

$K=\frac{Lwn}{r}$

\vspace{0.2cm}
with $w=1$, $L=1$, $n=12$, $r=0.10$, we obtain $K=120$. 
\vspace{0.2cm}

In real life a proportion of $120$ to $1$ between the productive capital per worker and the monthly compensation of a worker is not unrealistic.

\vspace{1cm}
\textbf{A comment}

This analytical framework allows us to account in a rigorous and non-partial way for the evolution of economic and social phenomena such as wealth and income distribution. In particular, our framework closely links distributional issues to the production process. Of course, this is a topic which has been extensively discussed in the history of economic thought and also in the public debate. \emph{E.g.}, \cite{galbraith2001distribution} discusses the policy implications of this linkage. 
In \cite{mazzoli_morini_terna_2019} this framework yet exists and is accounted for: in fact, the aggregate demand is explicitly formalized in a way that allows to account for income distribution, creating a link between distributional issues and technological process.

%%%%%%%%%%%%%%%%%%%%%%%%%%%%%%%%%%%%%%%%%%%%%%%%%%%%%%%%%%%%%%%%%%%%%%%%%%%%%%%%%%%%%%%
\section{Adapting the capital}\label{appAdapt}

\textbf{Adapting the quantity of the productive capital}

Scheme of the firm $i$ or $F_i$.

\vspace{0.2cm}
$L_t$ is the number of workers at time $t$.
\vspace{0.2cm}

$K_t$ is the productive capital in value at time $t$.
\vspace{0.2cm}

$K^q_t$ is the current capital in quantity, with $K^q_0  = K_0/p^K_0$ where $p^K_0$ is the initial price of capital goods;\footnote{The initial price is set internally by the program, estimating a value consistent with the costs of the firms producing goods for investments, considering also their markup. \textbf{Temporary}, the investment goods are supposed to be produced by a unique sector.} the updated mean price is indirectly calculated with $p^K_t=K_t/K^q_t$ as $\Delta K^q$ is determining $\Delta K$ using the updated capital goods price.
\vspace{0.2cm}

$\tau = $ tolerance. Within the $\pm$ tolerance interval we evaluate the current measure of the 
productive capital to be adequate. 
\vspace{0.2cm}

$K^q_{t_{min}} = \frac{K^q_t}{1+\tau}$ is the minimum value considered as adequate at time $t$.
\vspace{0.2cm}

$K^q_{t_{max}} = K^q_t (1+\tau)$ is the maximum value considered as adequate at time $t$.
\vspace{0.2cm}

$K^o_t = \frac{K_t}{u \cdot f}$ obsolescence and deterioration at time $t$ of the productive capital in value ($u=$ useful life of the productive capital and $f=$ time fraction in use) 
\vspace{0.2cm}

$K^{o^q}_t = \frac{K^q_t}{u \cdot f}$ obsolescence and deterioration at time $t$ of the productive capital in quantity ($u=$ useful life of the productive capital and $f=$ time fraction in use) 
\vspace{0.2cm}

The calculations of $K^o_t$ and $K^{o^q}_t$ are made independently considering the different prices adopted over time.

\vspace{0.2cm}

$K^{q^d}_{t+1}$ represents the desired capital in quantity at time $t+1$.
\vspace{0.2cm}

Looking back at time $t$, given an average quantity per order $\bar q_t$ and an average number $\bar n_t$ of orders, the firm is producing in parallel, and we have a reference quantity of products $Q_t = \bar q_t \bar n_t$.
\vspace{0.2cm}

With $l_p$ as labor productivity, at time $t+1$ we desire $L^d_{t+1} = Q_t/l_p$ as new labor quantity; with the recipe $\rho=K/L$, we desire $K^d_{t+1}=\rho L_{t+1}$; in quantity,  $K^{q^d}_{t+1}=K^d_{t+1}/p^K_{t}$. The correction of $L$ is made over a given interval (idiosyncratic of $F_i$); that of $K$ at each time unit. $p^K_{t}$ is employed as the best estimate of the future price of capital goods.

%\newpage

We have three cases:

\begin{itemize}
    
\item case I
\vspace{0.2cm}

$--K^{q^d}_{t+1}--$

$------< K^q_{t_{min}} ------- K^q_t -----K^q_{t_{max}} <------$

\item case II
\vspace{0.2cm}

~~~~~~~~~~~~~~~~~~~~~~~~~~~$-----K^{q^d}_{t+1}-------$

$------< K^q_{t_{min}} ------- K^q_t ------K^q_{t_{max}} <------$

\item case III
\vspace{0.2cm}

~~~~~~~~~~~~~~~~~~~~~~~~~~~~~~~~~~~~~~~~~~~~~~~~~~~~~~~~~~~~~~~~~~~~~~~~~~~~~~$--K^{q^d}_{t+1}--$

$------< K^q_{t_{min}} ------- K^q_t ------K^q_{t_{max}} <------$

\end{itemize}

In each case, the current values of $K_t$ and $K^q_t$ are decreasing of $K^o_t$ in value and of $K^{o^q}_t$ in quantity.
\vspace{0.2cm}

\vspace{0.2cm}

In case I, $K^{q^d}_{t+1} < K^q_{t_{min}}$:
\begin{itemize}
\item quantities:

\begin{enumerate}
\item $a=-K^{o^q}_t$ $\rightarrow$ reduction of the productive capital in quantity for obsolescence and deterioration at time $t$
\item $b=K^{q^d}_{t+1} - K^q_{t_{min}}$ (\emph{n.b.:} being $< 0$) $\rightarrow$ desired reduction for adaptation
\item if $b \le a$: $S=0$ $\rightarrow$ substitutions; if $b > a$: $S^q=|a| - |b|$ $\rightarrow$ substitutions 
\end{enumerate}

\item values:
\begin{enumerate}
\item $A=-K^{o}_t$ $\rightarrow$ reduction of the productive capital in value for obsolescence and deterioration at time $t$, using the implicit mean of the prices incorporated in all capital augmentations
\item $S=S^q \cdot p^K_t$ $\rightarrow$ substitutions in value
\end{enumerate}

\end{itemize}
\vspace{0.2cm}

In case II, $K^q_{t_{min}} \le K^{q^d}_{t+1} \le K^q_{t_{max}}$:
\begin{itemize}
\item quantities:
\begin{enumerate}
\item $a=-K^{o^q}_t$ $\rightarrow$ reduction of the productive capital in quantity for obsolescence and deterioration at time $t$
\item $S^q=|a|$ $\rightarrow$ substitutions
\end{enumerate}

\item values:
\begin{enumerate}
\item $A=-K^{o}_t$ $\rightarrow$ reduction of the productive capital in value for obsolescence and deterioration at time $t$, using the implicit mean of the prices incorporated in all capital augmentations
\item $S = S^q \cdot p^K_t$ $\rightarrow$ substitutions in value
\end{enumerate}

\end{itemize}
\vspace{0.2cm}

In case III, $K^q_{t_{max}} \le K^{q^d}_{t+1}$:

\begin{itemize}
\item quantities:
\begin{enumerate}
\item $a=-K^{o^q}_t$ $\rightarrow$ reduction of the productive capital in quantity for obsolescence and deterioration at time $t$
\item $S^q=|a|$ $\rightarrow$ substitutions
\item $\Delta^q= K^{q^d}_{t+1} - K^q_{t_{max}}$ $\rightarrow$ desired increment of the productive capital in quantity
\end{enumerate}

\item values:
\begin{enumerate}
\item $A=-K^{o}_t$ $\rightarrow$ reduction of the productive capital in value for obsolescence and deterioration at time $t$, using the implicit mean of the prices incorporated in all capital augmentations
\item $S=S^q\cdot p^K_t$ $\rightarrow$ substitutions in value
\item $\Delta=\Delta^q \cdot p^K_t$ $\rightarrow$ desired increment of the productive capital in value

\end{enumerate}

\end{itemize}

\end{appendices}

\end{document}